\documentclass[fleqn,usenatbib]{mnras}

\usepackage{newtxtext,newtxmath}

\usepackage[T1]{fontenc}

\DeclareRobustCommand{\VAN}[3]{#2}
\let\VANthebibliography\thebibliography
\def\thebibliography{\DeclareRobustCommand{\VAN}[3]{##3}\VANthebibliography}

\usepackage{graphicx}	
\usepackage{amsmath}	
\usepackage{subfigure}
\usepackage{paralist}
\usepackage{bm}
\usepackage{bbding}
\usepackage{multirow}

\title[Brown Hamiltonian]{Extensions of Brown Hamiltonian--I. A high-accuracy model for von Zeipel--Lidov--Kozai oscillations}

\author[Lei \& Grishin]{
Hanlun Lei$^{1,2}$\thanks{leihl@nju.edu.cn},
Evgeni Grishin$^{3,4}$
\\
$^{1}$School of Astronomy and Space Science, Nanjing University, Nanjing 210023, China\\
$^{2}$Key Laboratory of Modern Astronomy and Astrophysics in Ministry of Education, Nanjing University, Nanjing 210023, China\\
$^{3}$School of Physics and Astronomy, Monash University, Clayton, VIC 3800, Australia\\
$^{4}$OzGrav: Australian Research Council Centre of Excellence for Gravitational Wave Discovery, Clayton, VIC 3800, Australia
}

\date{Accepted XXX. Received YYY; in original form ZZZ}

\pubyear{2025}

\begin{document}
\label{firstpage}
\pagerange{\pageref{firstpage}--\pageref{lastpage}}
\maketitle

\begin{abstract}
Triple systems with low hierarchical structure are common throughout the Universe, including examples such as high-altitude lunar satellites influenced by the Earth, planetary satellites perturbed by the Sun, and stellar binaries affected by a supermassive black hole. In these systems, nonlinear perturbations are significant, making classical double-averaged models--even those incorporating the Brown Hamiltonian correction--insufficient for accurately capturing long-term dynamics. To overcome this limitation, the current study develops a high-precision dynamical model that incorporates the nonlinear effects of the quadrupole-order potential arising from both the inner and outer bodies, referred to as the extended Brown Hamiltonian model. This framework specifically expresses the Hamiltonian function and the transformation between mean and osculating orbital elements in elegant, closed forms with respect to the eccentricities of the inner and outer orbits. Practical applications to Jupiter’s irregular satellites show that the long-term evolutions predicted by the extended Brown Hamiltonian model align well with the results of direct $N$-body simulations. The developed Hamiltonian offers a fundamental dynamical model, which is particularly well suited for describing von Zeipel–Lidov–Kozai oscillations in low-hierarchy three-body systems.
\end{abstract}




\section{Introduction}
\label{Sect1}
In hierarchical triple systems, the eccentricity and inclination of the inner orbit undergo coupled evolution due to the long-term perturbation from a distant body. According to the classical theory, the perturbation from the third body can induce large oscillations in both eccentricity and inclination when the mutual inclination lies between $\sim$$39.2^{\circ}$ and $\sim$$140.8^{\circ}$ \citep{von1910application, kozai1962secular, lidov1962evolution}. This phenomenon is referred to as the von Zeipel–Lidov–Kozai (ZLK) oscillations \citep{ito2019lidov}. Applications of ZLK oscillations to varieties of astrophysical systems can be found in \citet{naoz2016eccentric} and \citet{shevchenko2016lidov}. 

Under the framework of the classical ZLK theory, a notable feature is that the condition of occurrence and the amplitude of excitation for ZLK oscillations are independent on the mass and distance of the perturber. This implies that a third body with an arbitrarily small mass and at an arbitrarily large separation can still excite similar ZLK oscillations of the inner binary \citep{tremaine2023hamiltonian}. However, the timescale (or period) of ZLK oscillation is closely related to the mass of perturber $m_p$ and its separation $a_{\rm p,eff}$ by \citep{antognini2015timescales}
\begin{equation}\label{Eq0}
t_{\rm ZLK}  \simeq  \frac{16}{15} \left(\frac{1}{n}\right) \left(\frac{m_0}{m_p}\right)\left(\frac{a_{\rm p,eff}}{a}\right)^3,
\end{equation}
where $a$ and $n$ are the semimajor axis and mean motion of the inner binary, $m_0$ is the mass of the central object, and $a_{\rm p,eff}$ is the perturber's effective semimajor axis defined by the semimajor axis $a_p$ and eccentricity $e_p$ in the form of $a_{\rm p,eff}= a_p \sqrt{1-e_p^2}$. The secular approximation requires that the timescales of a triple system are hierarchical, namely that $P_{\rm in} \ll P_{\rm out} \ll t_{\rm ZLK}$, where $P_{\rm in}$ and $P_{\rm out}$ are the Keplerian periods of the inner and outer binaries. In particular, short-period variations of orbital elements ($t \sim P_{\rm out}$) filtered out by the double averaging can be characterized by the single-averaging parameter \citep{luo2016double},
\begin{equation}\label{Eq01}
\epsilon_{\rm SA} = \left(\frac{a}{a_p \left(1-e_p^2\right)}\right)^{3/2} \frac{m_p}{\sqrt{m_0(m_0+m_p)}} \simeq \frac{8}{15\pi}\frac{P_{\rm out}}{t_{\rm ZLK}}
\end{equation}
which stands for the ratio of the outer binary period to ZLK oscillation timescale. The single-averaging parameter $\epsilon_{\rm SA}$ can be used to measure the hierarchy of triple systems \citep{luo2016double,liu2018black}.

When the single-averaging parameter $\epsilon_{\rm SA}$ is much smaller than unity, the timescale of ZLK oscillations is significantly longer than the Keplerian periods of both the inner and outer binaries, and the associated triple systems are characterized as highly hierarchical configurations. In such cases, the secular approximation or double-averaging approach (i.e., averaging over the orbits of both the inner and the outer bodies) can be employed to formulate what is known as the double-averaged Hamiltonian model \citep{broucke2003long,ford2000secular,naoz2016eccentric}.

As the single-averaging parameter $\epsilon_{\rm SA}$ increases, equation (\ref{Eq01}) indicates that the period (or timescale) of the ZLK oscillation decreases and thus the level of system hierarchy reduces, thereby challenging the validity of the secular and even quasi-secular approximations \citep{seto2013highly,liu2018black}.

In particular, when the indicator $\epsilon_{\rm SA}$ increases so that the period of the ZLK oscillation decreases to a level comparable to the Keplerian period of the outer binary, the corresponding triple systems are characterized as moderately hierarchical configurations, wherein the secular approximation may become inadequate. This limitation arises because the second-order perturbations of the quadrupole potential from the outer body begin to dominate the long-term evolution of the system. The second-order Hamiltonian for triple systems was first formulated and discussed by \citet{brown1936stellarI,brown1936stellarII,brown1936stellarIII,brown1937stellarIV}. For further discussion of Brown's work, please refer to \citet{cuk2004secular}. In \citet{tremaine2023hamiltonian}, the nonlinear Hamiltonian is referred to as Brown's Hamiltonian, in recognition of his pioneering contributions to this field. Historically, the difficulty in describing the secular behavior of distant satellites is most famously exemplified by Newton's inability to account for the precession of the lunar apsides \citep{brouwer1961methods,tremaine2023dynamics}. Brown's Hamiltonian, resulting from the modulation of the well-known evection terms in the third-body disturbing function, introduces a crucial correction to the lunar precession \citep{cuk2004secular,luo2016double,tremaine2023hamiltonian}. 

Brown's Hamiltonian has been extensively studied in the context of mildly hierarchical triple systems \citep{soderhjelm1975three,cuk2004secular,breiter2015secular,luo2016double,lei2018modified,krymolowski1999studies,will2021higher,li2025complete}. In the literature, there exist more than three different forms of Brown's Hamiltonian, which are related by the gauge freedom in canonical transformations \citep{tremaine2023hamiltonian}. Within the Brown Hamiltonian framework, modified maximal eccentricity, critical inclination, and fixed points have been analytically discussed by \citet{grishin2018quasi}, \citet{mangipudi2022extreme}, and \citet{grishin2024irregularI}. In the high-eccentricity regime, \citet{klein2024hierarchical} demonstrated that a significant effect of Brown's Hamiltonian on long-term evolution is to introduce an additional precession of the eccentricity vector. The quasi-secular theory based on Brown's Hamiltonian has been widely applied to understand Hill stability in arbitrarily inclined triple systems \citep{grishin2017generalized}, long-term evolution of irregular satellites in the Solar System \citep{cuk2004secular,grishin2024irregularI,grishin2024irregularII}, the formation of wide binaries in the Kuiper Belt \citep{grishin2020wide,rozner2020wide}, and compact object mergers along with gravitational waves \citep{fragione2019black,hamers2019double}.

Moreover, for extremely low hierarchy, the orbital and secular timescales are not well separated, and the ZLK oscillation timescale may be even comparable to the Keplerian period of the inner binary. Such systems are common in the Universe. Examples include high-altitude lunar satellites influenced by the Earth \citep{nie2019semi}, planetary satellites perturbed by the Sun \citep{cuk2004secular,beauge2006high,grishin2024irregularI,grishin2024irregularII,lei2019semi}, and stellar binaries affected by a supermassive black hole \citep{peissker2024binary,li2025complete}. In this regard, \citet{grishin2024irregularII} has noticed that the Brown Hamiltonian correction is inaccurate in describing the long-term dynamics of high-altitude retrograde satellites of giant planets. This is because, in low-hierarchy triple systems, the nonlinear effects arising from the inner orbit become significant, leading to the consequence that the single-averaging approximation may break down \citep{lei2018modified,liu2018black}. To address this issue, \citet{beauge2006high} and \citet{lei2019semi} employed elliptic expansions of the disturbing function to develop long-term dynamical models that incorporate nonlinear perturbations from both the inner and outer orbits. However, we notice that the expansion-based models have notable limitations. First, the complexity of these models makes them difficult to implement and apply in practice. Second, such models converge only for eccentricities below $e_c = 0.6627$ (the so-called Laplace limit), due to the use of Hansen coefficients in the elliptic expansions \citep{wintner1941analytical}. Third, because of its slow convergence, it is required to perform high-order expansion to reach a given precision in high-eccentricity regime, which is usually time consuming. These limitations largely restrict the applicability to real-world systems. To overcome this problem, it is expected to develop a simplified model that achieves precision comparable to the ones of \citet{beauge2006high} and \citet{lei2019semi}, while notably eliminating the need for elliptic expansions and enabling convergence at arbitrary eccentricities \citep{grishin2024irregularII}.

In this article, we arrive at the stated goal by developing a nonlinear long-term dynamical model, which we call the extended Brown Hamiltonian model. This framework elegantly provides both the Hamiltonian function and the transformation between mean and osculating orbital elements, expressed in closed forms concerning the eccentricities of the inner and outer orbits. Numerical applications to irregular satellites in the Solar System show that the secular evolution predicted by the extended Brown Hamiltonian model closely matches the results of direct $N$-body simulations. Further investigations of modified ZLK dynamics and practical applications will be presented in subsequent papers of this series.

The structure of this paper is organized as follows. In Section \ref{Sect2}, we briefly introduce the Hamiltonian function for hierarchical three-body systems. Section \ref{Sect3} develops the extended Brown Hamiltonian model. The transformation between mean and osculating elements is discussed in Section \ref{Sect4}. In Section \ref{Sect5}, we present several numerical examples with applications to Jupiter’s irregular satellites. Finally, Section \ref{Sect6} summarizes the conclusions of this work.

\section{Hamiltonian function}
\label{Sect2}
In this work, we consider a restricted hierarchical three-body system, where a test particle is moving around the central body with mass $m_0$ under the perturbation of a distant disturbing body with mass $m_p$. Under the test-particle approximation, the perturber's orbit around the central object remains stationary and corresponds to the invariable plane of the system. For convenience of description, we adopt an $m_0$-centered inertial coordinate system with the $z$-axis parallel to the angular momentum vector of the perturber's orbit $\bm{j}_p$, the $x$-axis pointing towards the perturber's eccentricity vector ${\bm e}_p$, and the $y$-axis completing the right-handed rule. The same reference frame is adopted by \citet{tremaine2023hamiltonian} in deriving Hamiltonian for ZLK oscillations.

Under the defined coordinate system, the motion of both the test particle and the perturber is described by the classical orbit elements, including the semimajor axis $a$, eccentricity $e$, inclination $i$, longitude of ascending node $\Omega$, argument of pericenter $\omega$ and mean anomaly $M$. Sometimes the mean anomaly $M$ is replaced by eccentric anomaly $E$ or true anomaly $f$. In our notations, the variables with subscript $p$ are for $m_p$ and the ones without any subscript are for the inner test particle. 

In a hierarchical configuration, the semimajor axis ratio $\alpha = a/a_p$ is a small parameter. As a result, the third-body disturbing function (per unit mass) can be expanded as a power series in $\alpha (=a/a_p)$ as follows \citep{harrington1969stellar}:
\begin{equation}\label{Eq1}
{\cal R} = \frac{{{\cal G}{m_p}}}{{{a_p}}}\sum\limits_{n = 2}^\infty  {{{\left( {\frac{a}{{{a_p}}}} \right)}^n}{{\left( {\frac{r}{a}} \right)}^n}} {\left( {\frac{{{a_p}}}{{{r_p}}}} \right)^{n + 1}}{P_n}\left( {\cos \psi } \right),
\end{equation}
where ${\cal G}$ is the universal gravitational constant, $\psi$ is the relative angle between the position vectors of the test particle and of the perturber relative to the central object (their distances are denoted by $r$ and $r_p$, respectively), and $P_n$ is the Legendre polynomial of order $n$. See Appendix \ref{A1} for the expressions of unitary position vectors of the perturber and of the test particle measured in the chosen coordinate system. 

For triple systems with low hierarchy, the stability condition requires that the semimajor axis ratio $\alpha$ is much smaller than unity, thus it is adequate to truncate the disturbing function at the quadrupole order,
\begin{equation}\label{Eq2}
{\cal R}_2 = \frac{{{\cal G}{m_p}}}{{{a_p}}}{\left( {\frac{a}{{{a_p}}}} \right)^2}{\left( {\frac{r}{a}} \right)^2}{\left( {\frac{{{a_p}}}{{{r_p}}}} \right)^3}\left( {\frac{3}{2}{{\cos }^2}\psi  - \frac{1}{2}} \right).
\end{equation}
In order to formulate an autonomous Hamiltonian model, we adopt the following augmented set of action--angle variables, i.e. Delaunay variables \citep{giacaglia1970semi,morbidelli2002modern},
\begin{equation}\label{Eq3}
\begin{aligned}
L =& \sqrt {\mu a} ,\quad G = L\sqrt {1 - {e^2}},\quad H = G\cos i,\quad {L_p},\\
l =& M,\quad\quad g = \omega,\quad\quad\quad\quad\;\; h = \Omega,\quad\quad\quad {l_p} = {M_p},
\end{aligned}
\end{equation}
where $\mu = {\cal G}m_0$ is the gravitational constant of the central body, and $L_p$ is the momentum conjugated to the perturber's mean anomaly $l_p$. Based on the canonical variables, the Hamiltonian function, governing the evolution of dynamical system, can be written as \citep{beauge2006high}
\begin{equation}\label{Eq4}
{\cal H} =  - \frac{{{\mu ^2}}}{{2{L^2}}} + {n_p}{L_p} - {\cal R}_2\left( {L,G,H,l,g,h,{l_p}} \right),
\end{equation}
where $n_p$ is the mean motion of the perturber. Generally, the third-body perturbation is much smaller than the two-body terms. Thus, the dynamical model can be treated by means of Hamiltonian perturbation theory \citep{Lei2024perturbation}, e.g. Lie-series transformation \citep{hori1966theory, deprit1969canonical} and von Zeipel transformation \citep{brouwer1959solution,kozai1962second,giacaglia1970semi}.

\begin{figure}
\centering
\includegraphics[width=\columnwidth]{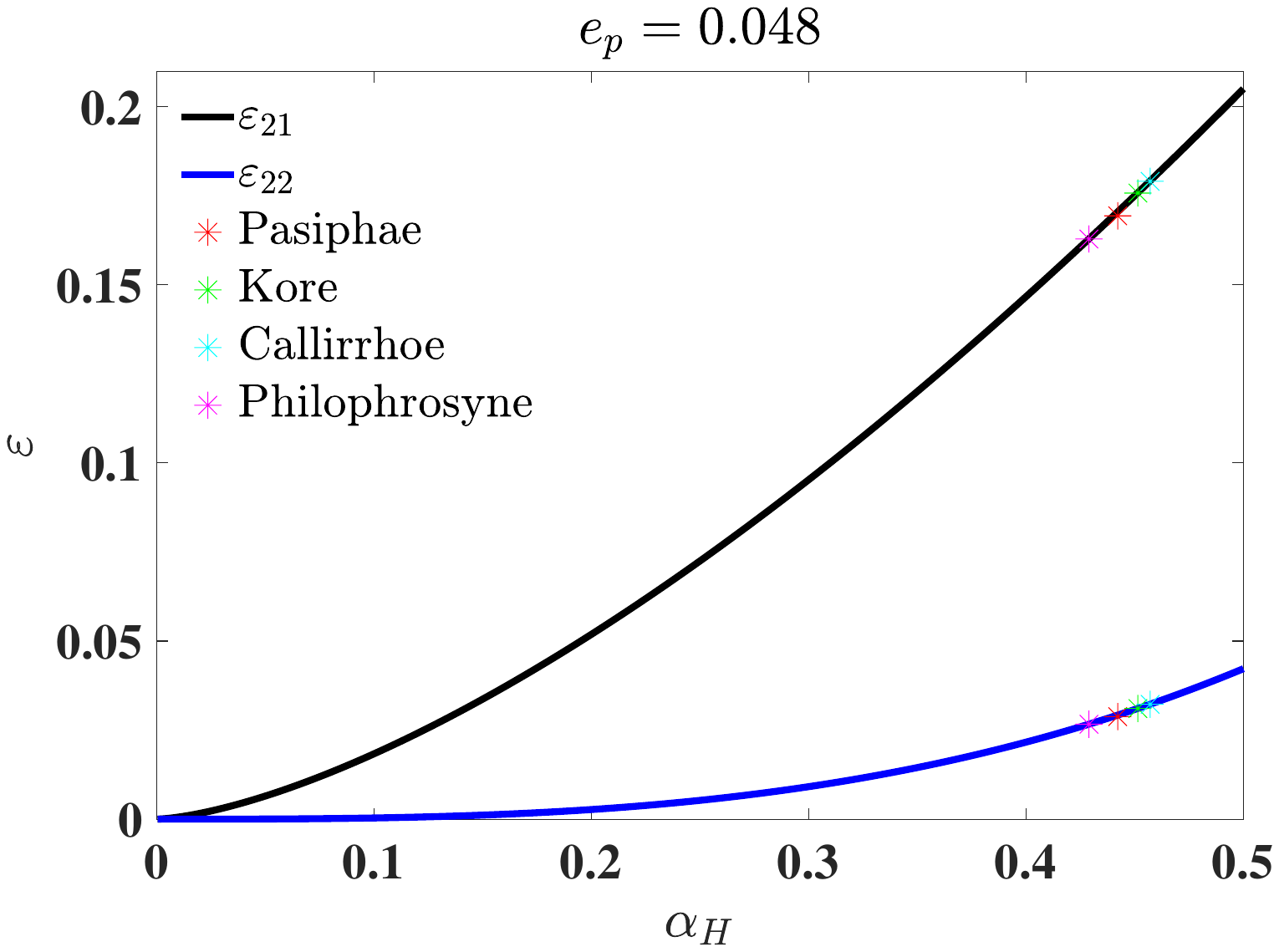}
\caption{Parameters $\varepsilon_{21}$ (measuring the contribution of classical Brown Hamiltonian) and $\varepsilon_{22}$ (measuring the contribution of the extended Brown Hamiltonian) are shown as functions of the ratio of the semimajor axis to the Hill radius $\alpha_H$ for the case of $e_p = 0.048$ (corresponding to the Jupiter--satellite--Sun system). The range of $\alpha_H$ is taken as $[0,0.5]$ for the sake of stability \citep{hamilton1991orbital}. The positions of four representative irregular satellites of Jupiter are marked in stars and they will be taken as examples for numerical simulations.}
\label{Fig1}
\end{figure}

\section{A closed form of the extended Brown Hamiltonian}
\label{Sect3}
In the Hamiltonian model, it is observed that those terms associated with angular coordinates $l$ and $l_p$ are of short period and their influences can be pushed to high orders by taking advantage of Hamiltonian perturbation technique. When we are focusing on test particles' long-term dynamics, high-order short-period effects can be filtered out by means of orbit averaging. In practice, we adopt von Zeipel method proposed by von Zeipel (1916) to handle this problem. Implementations of von Zeipel method in dealing with similar problems can be found in \citet{giacaglia1970semi}, \citet{naoz2013secular} and \citet{nie2019semi}, but nonlinear perturbations are not considered in these work.

The purpose of von Zeipel transformation is to realize a canonical transformation from a set of Delaunay variables to a new set of action--angle variables with the aid of generating function, in order to make the resulting Hamiltonian be independent on angular coordinates $l$ and $l_p$ \citep{brouwer1959solution}. See Appendix \ref{A2} for more details about the implementation of von Zeipel transformation. 

Up to the second order in canonical transformations, the closed form of the (normalized) disturbing function expressed in terms of eccentricity of the inner and outer orbits is composed of three parts,
\begin{equation}\label{Eq5}
{\cal F} = {{\cal F}_{20}} + {\varepsilon _{{\rm{21}}}}{{\cal F}_{21}} + {\varepsilon _{{\rm{22}}}}{{\cal F}_{22}},
\end{equation}
where ${\cal F}_{20}$ is the classical double-averaged quadrupole-order Hamiltonian term (widely appeared in literature), given by
\begin{equation}\label{Eq6}
{{\cal F}_{20}} = \frac{1}{6}\left( {2 + 3{e^2}} \right)\left( {3{{\cos }^2}i - 1} \right) + \frac{5}{2}{e^2}{\sin ^2}i\cos 2\omega,
\end{equation}
${\cal F}_{21}$ is the classical Brown Hamiltonian, given by 
\begin{equation}\label{Eq7}
{{\cal F}_{21}} = \frac{{3}}{{16}}\eta \cos i\left[ {2{{\sin }^2}i + {e^2}\left( {33 + 17{{\cos }^2}i + 15{{\sin }^2}i\cos 2\omega } \right)} \right],
\end{equation}
which represents the nonlinear effects of quadrupole-order perturbation, accumulated from those short-period oscillations associated with the outer binary, and ${\cal F}_{22}$ is termed \textbf{the extended Brown Hamiltonian},
\begin{equation}\label{Eq8}
\begin{aligned}
{{\cal F}_{22}} &= \frac{3}{{512}}\left\{ {227{e^2}\left( {8 + {e^2}} \right) - {{\cos }^4}i\left( {56 - 472{e^2} + 701{e^4}} \right)} \right.\\
&- 2{\cos ^2}i\left( {\frac{{376}}{3} - 360{e^2} - 305{e^4}} \right) - 95{e^4}{\sin ^4}i\cos 4\omega \\
&+ \left. {  4{e^2}{{\sin }^2}i\left[ {186 + \frac{{109}}{3}{e^2} + 5\left( {18 - 37{e^2}} \right){{\cos }^2}i} \right]\cos 2\omega } \right\},
\end{aligned}
\end{equation}
which stands for the nonlinear effects of quadrupole-order perturbation, accumulated from those short-period oscillations associated with the inner binary. In equation (\ref{Eq7}), we denote $\eta = \sqrt{1-e^2}$. See Appendix \ref{A2} for more details about the implementation of von Zeipel transformation and the correspondence of ${\cal F}_{20}$, ${\cal F}_{21}$ and ${\cal F}_{22}$. The vectorial form of the extended Brown Hamiltonian model is presented in Appendix \ref{A5}. For convenience of comparison, the complete Hamiltonian including the octupole-order term is provided in Appendix \ref{A3}.

The terminology `Brown Hamiltonian' (often called second-order Hamiltonian) first appeared in \citet{tremaine2023hamiltonian}. The classical Brown Hamiltonian is discussed in \citet{brown1936stellarIII}, and there are more than three different versions of the Brown Hamiltonian in the literature \citep{cuk2004secular,luo2016double,breiter2015secular,lei2018modified,soderhjelm1975three,will2021higher,tremaine2023hamiltonian,krymolowski1999studies}. Recently, \citet{tremaine2023hamiltonian} demonstrated that these various expressions of the Brown Hamiltonian are related by the gauge freedom in canonical transformations, reflecting the dependence of the Brown Hamiltonian on the choice of fictitious time. Notably, the Brown Hamiltonian ${{\cal F}_{21}}$ presented in equation (\ref{Eq7}) corresponds to the simplest form and is consistent with equation (64) of \citet{tremaine2023hamiltonian}. 

In equation (\ref{Eq5}), the significance of the classical and extended Brown Hamiltonian terms are measured by the coefficients $\varepsilon_{21}$ and $\varepsilon_{22}$, given by
\begin{equation}\label{Eq9}
{\varepsilon _{21}} = \left( {\frac{{{n_p}}}{n}} \right)\left( {\frac{{{m_p}}}{{{m_0} + {m_p}}}} \right)\frac{1}{{{{\left( {1 - e_p^2} \right)}^{3/2}}}}\left( {1 + \frac{2}{3}e_p^2} \right),
\end{equation}
and
\begin{equation}\label{Eq10}
{\varepsilon _{22}} = {\left( {\frac{{{n_p}}}{n}} \right)^2}\left( {\frac{{{m_p}}}{{{m_0} + {m_p}}}} \right)\frac{1}{{{{\left( {1 - e_p^2} \right)}^3}}}\left( {1 + 3e_p^2 + \frac{3}{8}e_p^4} \right).
\end{equation}
The coefficient $\varepsilon_{21}$ is related to the single-averaging parameter $\epsilon_{\rm SA}$ by
\begin{equation}\label{Eq10-1}
\varepsilon_{21} = \epsilon_{\rm SA} \left( {1 + \frac{2}{3}e_p^2} \right),
\end{equation}
where $\epsilon_{\rm SA}$ measures the magnitude of short-period oscillations of orbital elements associated with the outer binary period \citep{luo2016double}. It should be noted that the expression of $\varepsilon_{21}$ can be summarized from \citet{tremaine2023hamiltonian} and \citet{klein2024hierarchical}, while $\varepsilon_{22}$ is a new indicator measuring the short-period variations associated with the inner binary period.

In the case of $m_0 \ll m_p$ (e.g., irregular satellites of giant planets perturbed by the Sun), Hill radius of the central object can be approximated by \citep{tremaine2023dynamics}
\begin{equation}\label{Eq10-2}
r_H = a_p \left(\frac{m_0}{3 m_p}\right)^{1/3},
\end{equation}
and thus the coefficients can reduce to
\begin{equation}\label{Eq10-3}
\begin{aligned}
{\varepsilon _{21}} &= \frac{1}{{\sqrt 3 }}{\left( {\frac{a}{{{r_H}}}} \right)^{3/2}}\frac{1}{{{{\left( {1 - e_p^2} \right)}^{3/2}}}}\left( {1 + \frac{2}{3}e_p^2} \right),\\
{\varepsilon _{22}} &= \frac{1}{3}{\left( {\frac{a}{{{r_H}}}} \right)^3}\frac{1}{{{{\left( {1 - e_p^2} \right)}^3}}}\left( {1 + 3e_p^2 + \frac{3}{8}e_p^4} \right).
\end{aligned}
\end{equation}
It is observed that the factors $\varepsilon_{21}$ and $\varepsilon_{22}$ are dependent on the ratio of semimajor axis to the Hill radius $\alpha_H = a/r_H$ as well as the eccentricity of the perturber's orbit $e_p$, indicating that the curves of $\varepsilon_{21}$ and $\varepsilon_{22}$ in the space of $\alpha_H$ are independent on mass parameters. 

Furthermore, for the case of nearly circular-orbit perturbation, the coefficients can further reduce to
\begin{equation}\label{Eq11-1}
{\varepsilon _{21}} = \frac{1}{{\sqrt 3 }}{\alpha_H^{3/2}} \sim \frac{8}{15\pi} \frac{P_{\rm out}}{t_{\rm ZLK}},\quad {\varepsilon _{22}} = \frac{1}{3}{\alpha_H^3} \sim \frac{8}{15\pi} \frac{P_{\rm in}}{t_{\rm ZLK}},
\end{equation}
where ${\varepsilon _{21}}$ is equivalent to the single-averaging parameter $\epsilon_{\rm SA}$. We can see that the coefficients $\varepsilon_{21}$ and $\varepsilon_{22}$ possess clear physical significance, as they quantify the ratio of the outer and inner binary periods to the ZLK oscillation timescale, respectively.

Figure \ref{Fig1} shows the distributions of $\varepsilon_{21}$ and $\varepsilon_{22}$ as functions of $\alpha_H$ for the case of $e_p = 0.048$, which corresponds to the Jupiter--satellite--Sun system. To ensure the stability of dynamical system, the upper limit of $\alpha_H$ is fixed at 0.5 \citep{hamilton1991orbital}\footnote{Highly inclined satellites could have a much lower Hill stability limit of around $\sim 0.4$ due to high eccentricity raised by the ZLK effect \citep{grishin2017generalized}.}.
It is observed that both $\varepsilon_{21}$ and $\varepsilon_{22}$ are increasing functions of $\alpha_H$, indicating that both the classical and extended Brown Hamiltonian terms have greater contributions for test particles moving on higher-altitude orbits.

Some discussions are made about the extended Hamiltonian model. 
\begin{itemize}
    \item The Hamiltonian specified by equation (\ref{Eq5}) determines an elegant and integrable dynamical model, which is of single degree of freedom with $(g,G)$ as the unique pair of phase-space variables. Similar to the classical ZLK oscillation discussed in \citet{kinoshita1999analytical} and \citet{kinoshita2007general}, the trajectories under the extended Brown Hamiltonian model can be expressed analytically as Jacobian elliptic functions. 
    \item The actions $L$ and $H$ are motion integral because the angular coordinates $l$ and $h$ are both cyclic coordinates, indicating that the orbit energy of the inner binary conserves and the eccentricity exchanges with inclination during long-term evolutions. 
    \item The deviation to the classical ZLK Hamiltonian model (specified by ${\cal F}_{20}$) is controlled by the small factors $\varepsilon_{21}$ and $\varepsilon_{22}$. It means that the approximate solutions of the extended Brown Hamiltonian model can be formulated by taking the solution of ZLK Hamiltonian as a starting point. This issue will be discussed in the next paper of this series.
    \item The classical Brown Hamiltonian is of first order in mean motion ratio between the inner and outer binaries ($n_p/n$), and the extended Brown Hamiltonian is of second order in mean motion ratio. It shows that the extended Brown Hamiltonian ${\cal F}_{22}$ should be taken into account for those low-hierarchy three-body systems where the mean motion ratio ($n_p/n$) is not small.
\end{itemize}

In Section \ref{Sect5}, we will examine three long-term Hamiltonian models with increasing levels of accuracy: a) the standard double-averaged model (${\cal F}_{20}$ only), b) the brown Hamiltonian model (turning on $\varepsilon_{21}$), and c) our novel extended model (turning on both $\varepsilon_{21}$ and $\varepsilon_{22}$). The equations of motion under the extended Brown Hamiltonian model are provided in Appendix \ref{A4}.

\section{A closed form of transformation between mean and osculating elements}
\label{Sect4}

According to von Zeipel transformation, the generating function can be utilized to provide a transformation between the old and new sets of Delaunay variables. Here, we provide the transformation in terms of orbital elements, consistent with the generating function $S$ in Appendix \ref{A2}. It should be mentioned that an equivalent transformation can be reached on the basis of Delaunay variables.

For convenience of description, we refer to the variables under the original Hamiltonian model governed by ${\cal H}$ as osculating elements, denoted by $(a,e,i,\Omega,\omega,M)$, and refer to the ones under the extended Brown Hamiltonian model governed by ${\cal F}$ as mean elements, denoted by $(a^*,e^*,i^*,\Omega^*,\omega^*,M^*)$. Similar notations can be found in \citet{lei2019semi}. The transformation between mean and osculating elements is expressed by
\begin{equation}\label{Eq12}
\begin{aligned}
a =& {a^*} + \delta a,\quad\;\; e = {e^*} + \delta e,\quad\quad i = {i^*} + \delta i,\\
\Omega  =& {\Omega ^*} + \delta \Omega,\quad \omega  = {\omega ^*} + \delta \omega,\quad M = {M^*} + \delta M,
\end{aligned}
\end{equation}
where the periodic oscillations associated with the inner and outer orbits can be obtained by combining the generating function $S$ with the Lagrange planetary equations \citep{murray1999solar,tremaine2023dynamics},
\begin{equation}\label{Eq13}
\begin{aligned}
\delta a &= \frac{2}{{na}}\frac{{\partial S}}{{\partial M}},\quad \delta e = \frac{\eta }{{n{a^2}e}}\left( {\eta \frac{{\partial S}}{{\partial M}} - \frac{{\partial S}}{{\partial \omega }}} \right),\\
\delta i &= \frac{1}{{n{a^2}\eta }}\frac{1}{{\sin i}}\left( {\cos i\frac{{\partial S}}{{\partial \omega }} - \frac{{\partial S}}{{\partial \Omega }}} \right),\quad \delta \Omega  = \frac{1}{{n{a^2}\eta }}\frac{1}{{\sin i}}\frac{{\partial S}}{{\partial i}},\\
\delta \omega  &= \frac{\eta }{{n{a^2}}}\left( {\frac{1}{e}\frac{{\partial S}}{{\partial e}} - \frac{{\cot i}}{{{\eta ^2}}}\frac{{\partial S}}{{\partial i}}} \right),\quad \delta M =  - \frac{2}{{na}}\frac{{\partial S}}{{\partial a}} - \frac{{{\eta ^2}}}{{n{a^2}e}}\frac{{\partial S}}{{\partial e}}
\end{aligned}
\end{equation}
with $\eta=\sqrt{1-e^2}$. The generating function $S$ is provided by equation (\ref{EqA20}). It should be emphasized that the functions on the right-hand side of equation (\ref{Eq13}) are evaluated at mean elements $(a^*,e^*,i^*,\Omega^*,\omega^*,M^*)$.  

Because the expression of generating function $S$ shown by equation (\ref{EqA20}) is convergent at arbitrary eccentricities, we can state that a closed form is reached for the transformation between mean and osculating elements. In addition, for those low-eccentricity configurations, we have numerically checked that the closed-form transformation developed in this work is exactly equivalent to that discussed in \citet{lei2019semi}, where the transformation is realized based on elliptic expansions of the disturbing function.

\begin{table}
\begin{center}	
\caption{Parameters of the irregular satellites considered in Figs. \ref{Fig2}--\ref{Fig4}, including the periods of inner and outer binaries, timescale of ZLK oscillations and the coefficients $\varepsilon_{21}$ and $\varepsilon_{22}$.} \label{Table0}
{\begin{tabular}{l c c c c c}\hline\hline
Object&$P_{\rm in}$ (yr)&$P_{\rm out}$ (yr)&$t_{\rm ZLK}$ (yr)&$\varepsilon_{21}$&$\varepsilon_{22}$\\
\hline
Pasiphae&1.999&11.859&11.911&0.169&0.029\\
Kore&2.077&11.859&11.468&0.176&0.031\\
Callirrhoe&2.115&11.859&11.258&0.179&0.032\\
Philophrosyne&1.923&11.859&12.381&0.163&0.027\\
\hline
\end{tabular}}
\end{center}
\end{table}

\section{Numerical simulations}
\label{Sect5}

In this section, long-term dynamical models formulated in Section \ref{Sect3} are applied to long-term predictions of test particles under low-hierarchical triple systems, with consideration of mean-to-osculating transformations developed in Section \ref{Sect4}. In practice, we take irregular satellites of giant planets as representative examples of low hierarchies. Irregular satellites are typically defined as those that orbit the giant planets well beyond major moons, often exhibiting highly eccentric and inclined orbits \citep{cuk2004secular}. Regarding irregular satellites, the double-averaged models (even with Brown correction) are inadequate to predict their long-term behaviors \citep{carruba2002inclination,cuk2004secular,beauge2006high,lei2018modified,lei2019semi,grishin2024irregularI,grishin2024irregularII}.    

\subsection{Numerical examples}
\label{Sect5-1}

Four representatives of Jupiter's irregular satellites are taken as examples, including Pasiphae, Kore, Callirrhoe and Philophrosyne. Their parameters are shown in Table \ref{Table0}, including the periods of inner and outer binaries, timescale of ZLK oscillations and the coefficients $\varepsilon_{21}$ and $\varepsilon_{22}$. It is observed that the ZLK oscillation timescale is comparable to the outer binary period and even to the level of inner binary period.

\begin{figure*}
\centering
\includegraphics[width=\columnwidth]{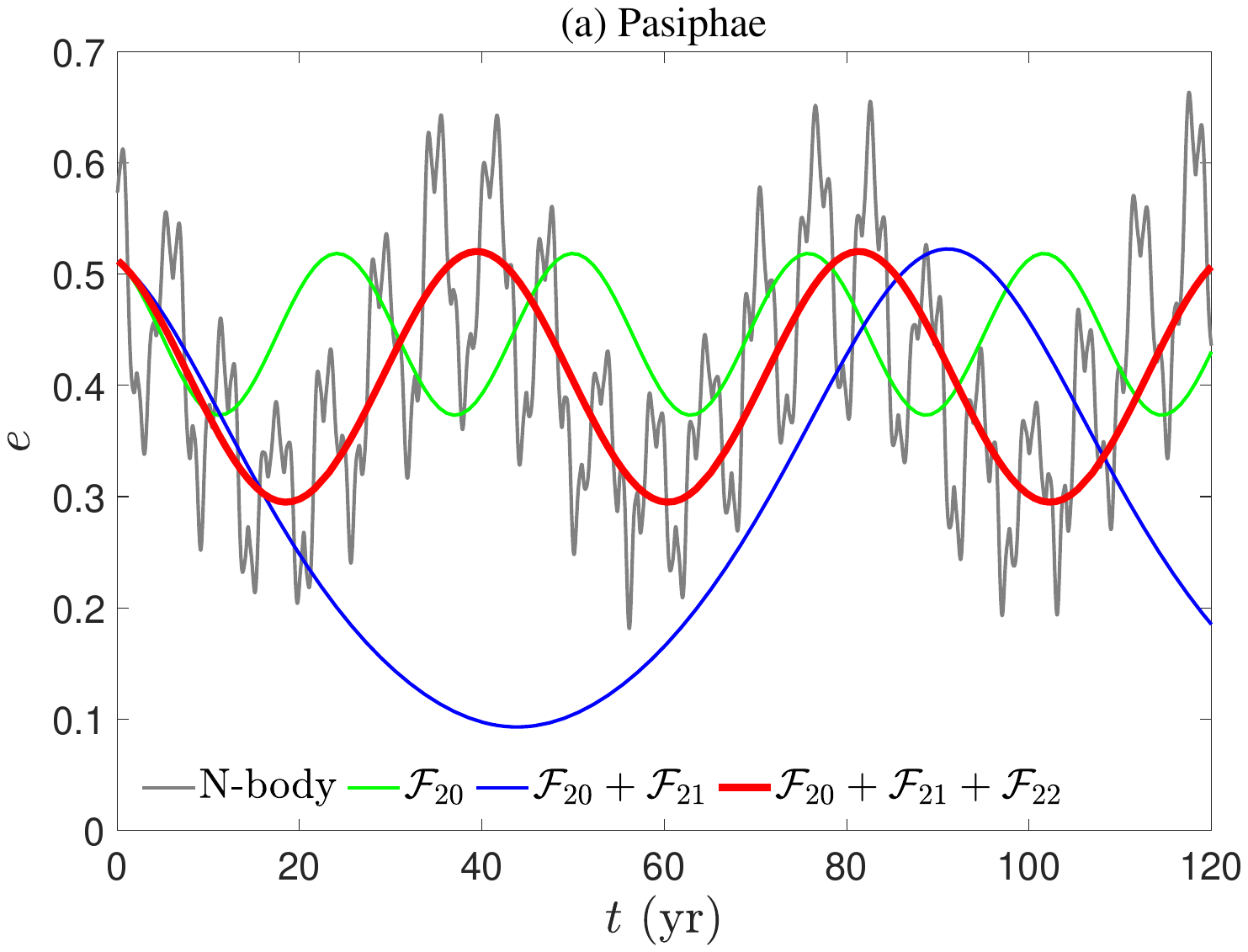}
\includegraphics[width=\columnwidth]{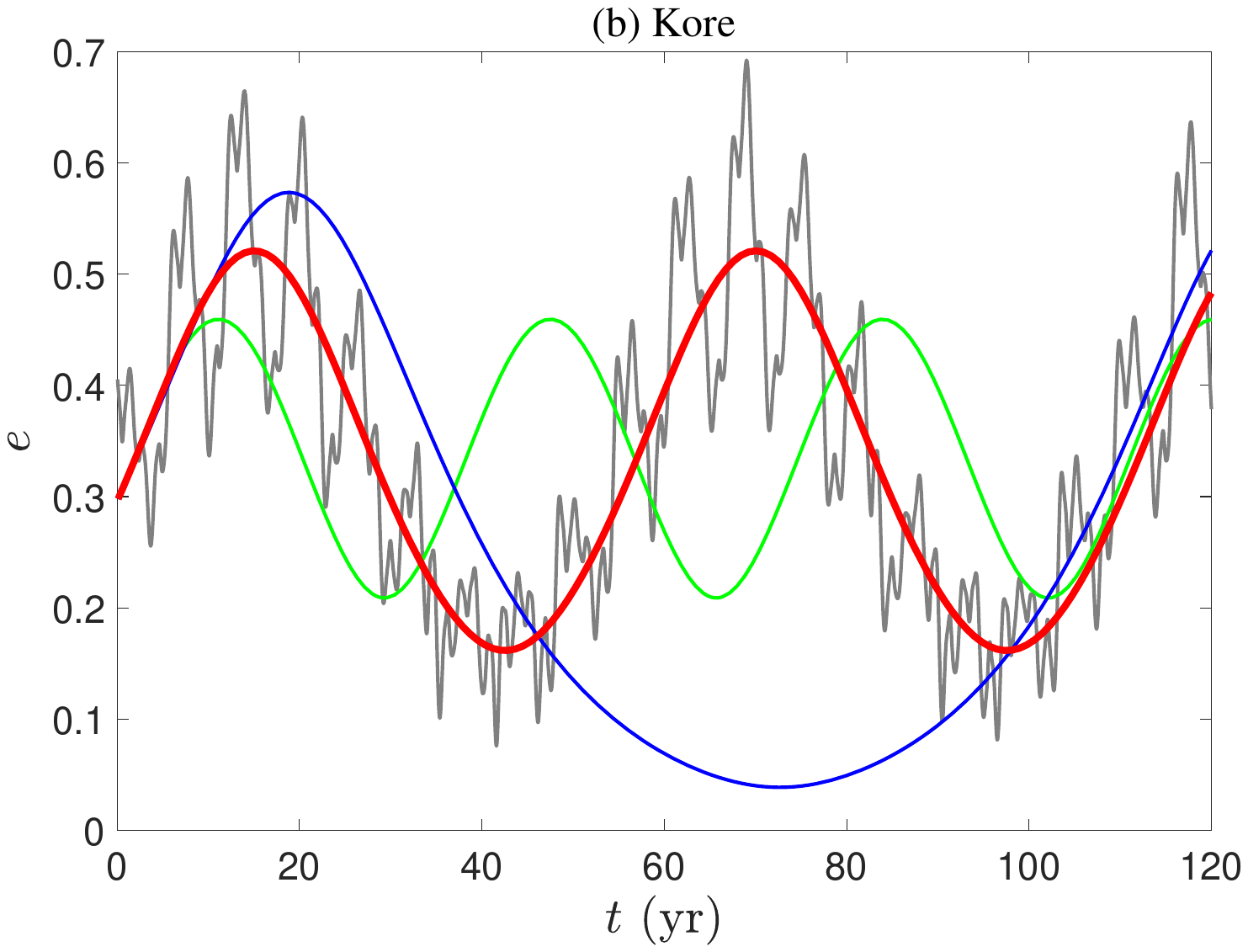}
\caption{Time histories of eccentricity for Jupiter's irregular satellites Pasiphae (\textit{left panel}) and Kore (\textit{right panel}), propagated under the full $N$-body model (see grey lines) and three long-term dynamical models, including the classical double-averaged model denoted by ${\cal F}_{20}$ (see green lines), the classical Brown Hamiltonian model denoted by ${\cal F}_{20} + {\cal F}_{21}$ (see blue lines), and the extended Brown Hamiltonian model denoted by ${\cal F}_{20} + {\cal F}_{21}+ {\cal F}_{22}$ (see red lines). In our notations, ${\cal F}_{20}$ stands for the classical quadrupole-order term, ${\cal F}_{21}$ for the classical Brown term and ${\cal F}_{22}$ for the extended Brown term.}
\label{Fig2}
\end{figure*}

\begin{figure*}
\centering
\includegraphics[width=\columnwidth]{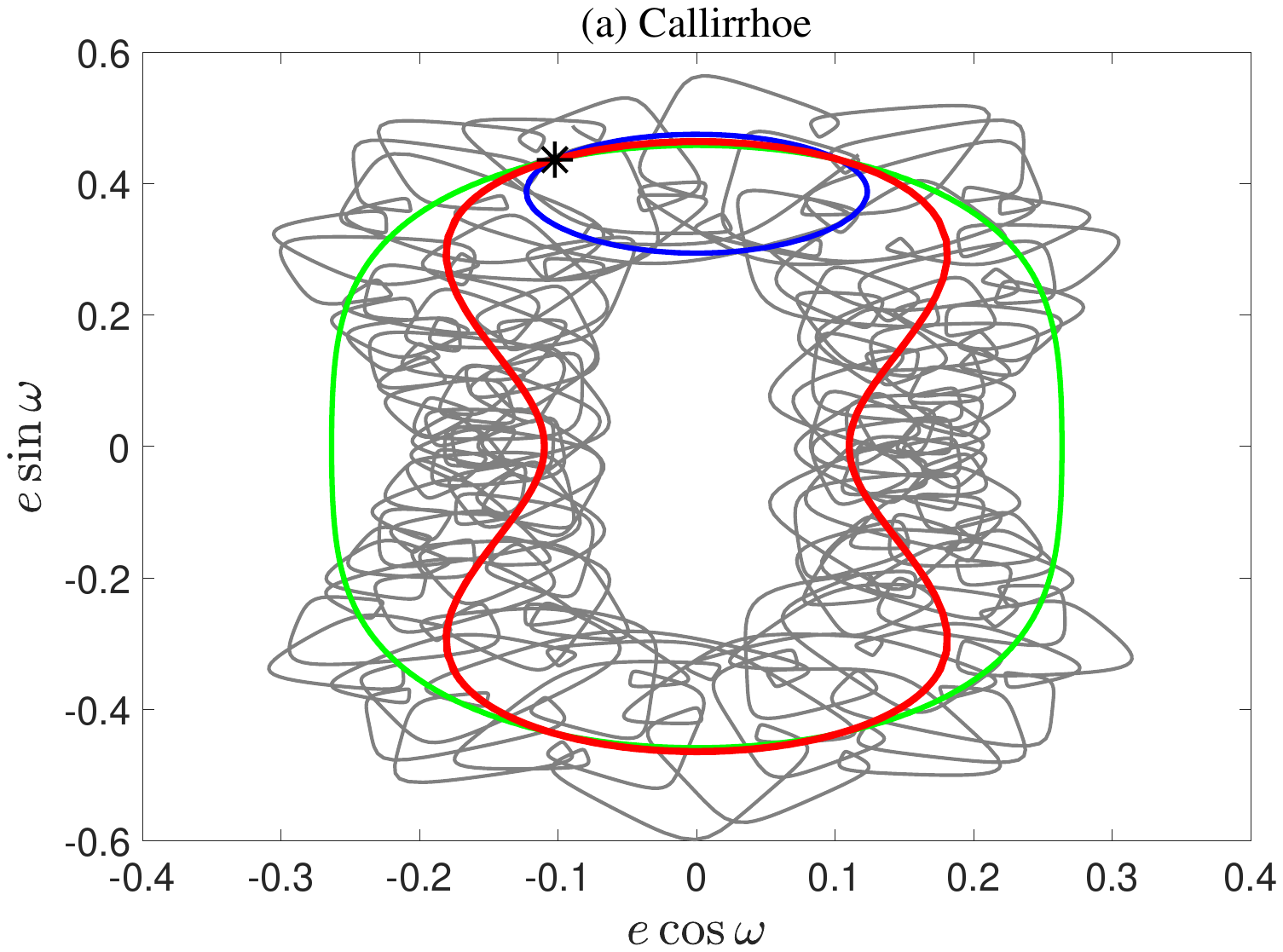}
\includegraphics[width=\columnwidth]{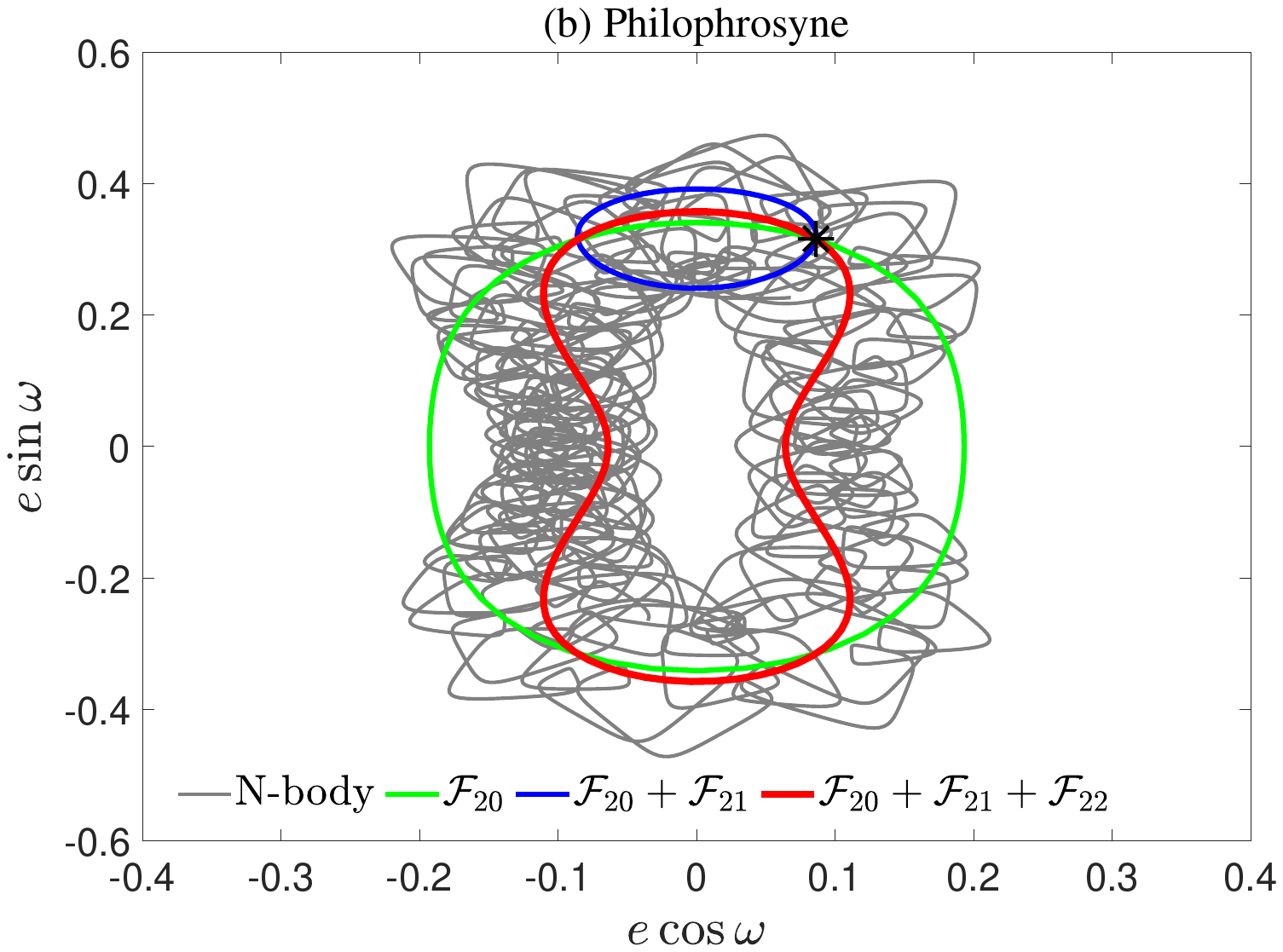}
\caption{Dynamical evolutions in the space $(e\cos{\omega}, e\sin{\omega})$ for Jupiter's irregular satellites including Callirrhoe (\textit{left panel}) and Philophrosyne (\textit{right panel}), propagated under different dynamical models. In both panels, the black stars stand for the starting points for trajectories propagated under long-term dynamical models.}
\label{Fig3}
\end{figure*}

Long-term behaviors are produced under three Hamiltonian models including ${\cal F}_{20}$, ${\cal F}_{20} + {\cal F}_{21}$ and ${\cal F}_{20} + {\cal F}_{21}+ {\cal F}_{22}$ models, and they are compared to the osculating trajectories propagated under the full $N$-body model\footnote{We employed a classic eighth-order Runge–Kutta numerical integrator, which utilizes a seventh-order method for automatic step-size control \citep{fehlberg1969classical}. The relative tolerance was set as $1 \times 10^{-14}$ to reach high precision}. Table \ref{Table1} provides the mean and osculating elements\footnote{Taken from https://ssd.jpl.nasa.gov/horizons/ at epoch October 18, 2024.} used for initial conditions of long-term Hamiltonian models and $N$-body integrations, respectively. Figure \ref{Fig2} shows the results about Pasiphae and Kore, and Fig. \ref{Fig3} presents the results for Callirrhoe and Philophrosyne. Numerical results (circulation of $\omega$) show that all the chosen examples are outside the ZLK resonance.

From Fig. \ref{Fig2}, it is observed that, for Pasiphae and Kore, the outcome of the direct $N$-body integration can be excellently reproduced by the extended Brown Hamiltonian model (${\cal F}_{20} + {\cal F}_{21}+ {\cal F}_{22}$). However, there are significant deviations in terms of the period and amplitude of ZLK oscillations for both the classical double-averaged model (${\cal F}_{20}$) and the classical Brown Hamiltonian model (${\cal F}_{20} + {\cal F}_{21}$). In particular, the period of ZLK oscillation is reduced under the ${\cal F}_{20}$ model, while it is enlarged under the ${\cal F}_{20} + {\cal F}_{21}$ model.

For Callirrhoe and Philophrosyne, Fig. \ref{Fig3} shows that the outcome of direct $N$-body integration can be perfectly reproduced by the extended Brown Hamiltonian model. Similarly, significant deviations can be observed under both the ${\cal F}_{20}$ and ${\cal F}_{20} + {\cal F}_{21}$ models. In particular, the ${\cal F}_{20} + {\cal F}_{21}$ model predicts that Callirrhoe and Philophrosyne are both inside the ZLK resonance ($\omega$ librates around a center within a limited range), which is an opposite (and wrong) prediction.

Direct comparisons made in Figs. \ref{Fig2} and \ref{Fig3} lead us to a firm conclusion that the extended Brown Hamiltonian term ${\cal F}_{22}$ has significant contributions to long-term evolutions\footnote{Here long term means that the evolution of orbital element is considered over the timescale of ZLK oscillation.}. In practice, such a novel term should be taken into account in order to describe ZLK oscillations accurately, especially for those low-hierarchy three-body systems where the mean motion ratio is not small.

\begin{figure}
\centering
\includegraphics[width=\columnwidth]{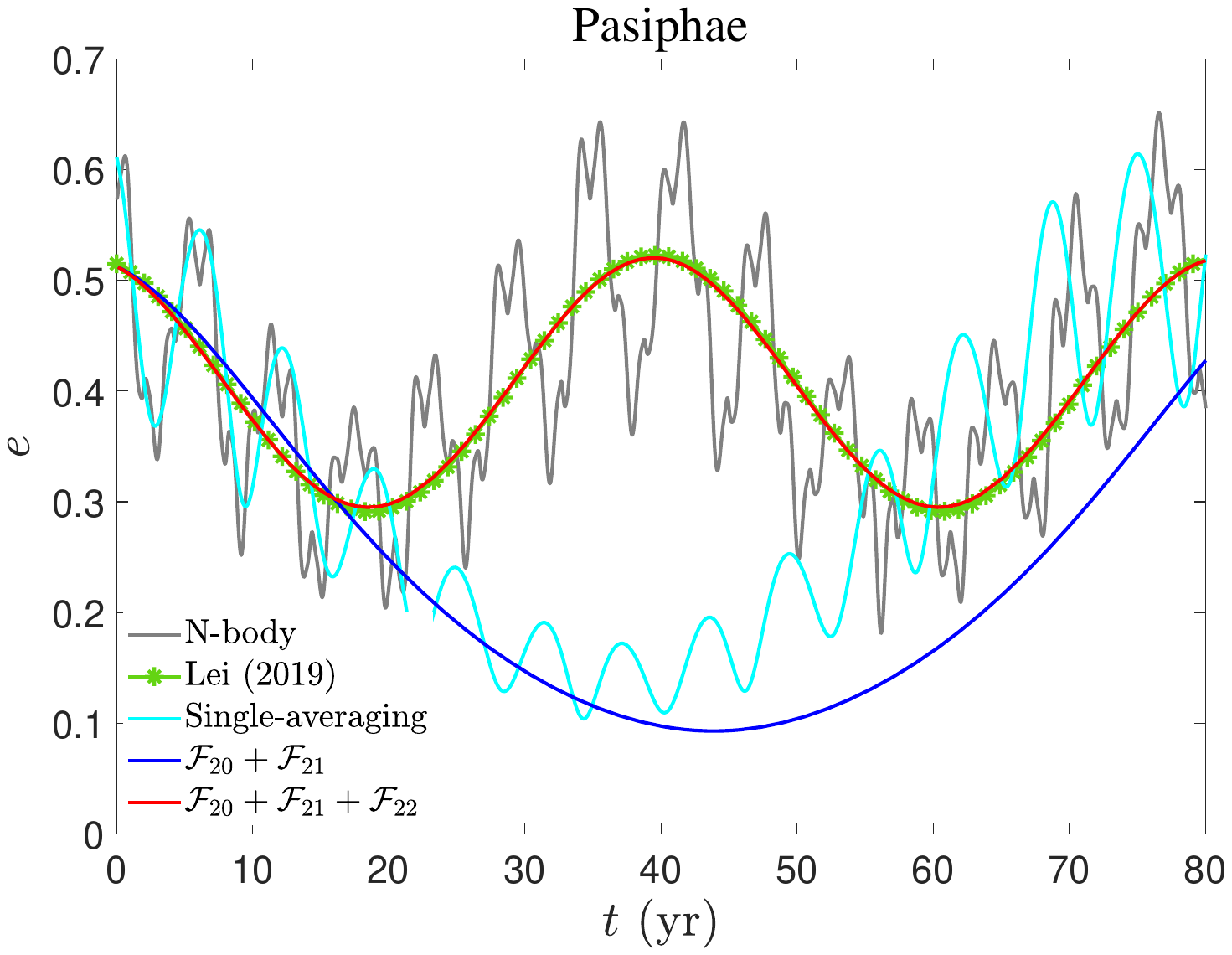}
\caption{Comparison of Pasiphae's eccentricity evolutions between the Hamiltonian model developed in \citet{lei2019semi}, the single-averaged model governed by the Hamiltonian of equation (\ref{EqA9-1}), and the classical and extended Brown Hamiltonian models developed in this work. The trajectories propagated under the full $N$-body model are shown in grey lines as background.}
\label{Fig4}
\end{figure}

\subsection{Discussions}
\label{Sect5-2}

The Hamiltonian given by equation (\ref{EqA9-1}) determines the single-averaged dynamical model, where only the average over the inner binary period is performed for the Hamiltonian function. The generating function $S_1$ presented by equation (\ref{EqA21}) provides the transformation between the osculating elements and the associated `single-averaged' elements. Figure \ref{Fig4} shows a comparison of Pasiphae's evolutions of eccentricity between the single-averaged model and the (classical and extended) Brown Hamiltonian models, together with the $N$-body simulations as the background. It is observed that there are significant discrepancy between the $N$-body outcome and the results of single-averaged model, showing that those short-period terms in the timescale of $P_{\rm in}$ (period of inner binary) produce significant influences upon the ZLK oscillations. This may help us to understand the reason that the classical Brown Hamiltonian model (${\cal F}_{20} + {\cal F}_{21}$) fails to predict the long-term behaviors, as observed in Figs. \ref{Fig2} and \ref{Fig3}.

It is noted that the same topic of this paper has been considered in \citet{lei2019semi} from the viewpoint of elliptic expansion of the disturbing function. Figure \ref{Fig4} also compares the long-term evolutions of Pasiphae produced from the extended Brown Hamiltonian model developed in this work and from the model formulated in \citet{lei2019semi}. Good agreement can be observed between them. The slight deviation between them may be due to the truncation of elliptic expansions. 

Due to the complexity of elliptic expansion, the Hamiltonian model developed in \citet{lei2019semi} is not easy for readers in handling practical problems. In addition, it is known that the eccentricity should be smaller than $e_c = 0.6627$ to ensure the convergence of elliptic expansions. Thus, its applicability is strictly restricted by the Laplace limit. When the eccentricity increases, the disturbing function needs to be expanded up to an extremely high order in order to reach a given precision. Such an expansion process is usually time-consuming, further limiting its applications. However, the extended Brown Hamiltonian developed in this work has an elegant expression, and thus the corresponding model is computationally friendly. In addition, the developed Hamiltonian has a closed form in terms of eccentricities of the inner and outer binaries, showing that there is no restriction for eccentricities in practical applications. 

\begin{table*}
\begin{center}	
\caption{Osculating and mean elements of irregular satellites used for providing initial conditions of the trajectories shown in Figs. \ref{Fig2}--\ref{Fig4}. Under the Jupiter-centered coordinate system, the orbit elements of the Sun (acting as a perturber) moving around the Jupiter are $a_{p} = 5.2018958475 (\rm au)$, $e_{p} = 0.0482582593$, $M_{p} = 52.9543354023^{\circ}$, and the remaining orbit elements are not defined and thus they can be arbitrarily taken. In our notations, `Oscu.' represents osculating elements and `Mean' stands for mean elements.} \label{Table1}
{\begin{tabular}{l l c c c c c c c}\hline\hline
Object&Type&$a (\rm au)$&$e$&$i(^{\circ})$&$\Omega(^{\circ})$&$\omega(^{\circ})$&$M(^{\circ})$& Figs.\\
\hline
\multirow{2}{*}{Pasiphae}&Oscu.&0.1569589992&0.5752369565&154.9133780476&236.1548433354&277.4076490881&352.2028055842&\multirow{2}{*}{\ref{Fig2}(a) \& \ref{Fig4}}\\
&Mean&0.1562598702&0.5126090410&153.4501837809&235.8880572761&284.0439665656&344.6601941803&\\
\hline
\multirow{2}{*}{Kore}&Oscu.&0.1596079995&0.4101328535&141.7700642470&207.4929063833&230.9114514631&266.2084740484&\multirow{2}{*}{\ref{Fig2}(b)}\\
&Mean&0.1602618820&0.2959616683&140.1994261380&209.6168345516&224.4194603433&275.3879232429&\\
\hline
\multirow{2}{*}{Callirrhoe}&Oscu.&0.1523808990&0.4661923617&149.3948821000&186.4719027703&75.1742046169&136.2867190150&\multirow{2}{*}{\ref{Fig3}(a)}\\
&Mean&0.1622528053&0.4492399178&148.2796232101&195.6101050159&103.0714660907&110.4454014826&\\
\hline
\multirow{2}{*}{Philophrosyne}&Oscu.&0.1474758882&0.2389517050&146.1071565289&123.8051056218&72.5537063231& 62.6458179972&\multirow{2}{*}{\ref{Fig3}(b)}\\
&Mean&0.1522819182&0.3270333831&148.0457019681&121.4474749588&74.7619194183&55.7959265236&\\
\hline
\end{tabular}}
\end{center}
\end{table*}

\section{Conclusions}
\label{Sect6}

The main contribution of this paper is to develop the next order of approximation for modeling triple systems of low hierarchies, where even the Brown Hamiltonian correction is not accurate enough to describe the long-term evolution.

Firstly, we arrived at an elegant expression of the long-term Hamiltonian, presented in closed form in terms of the eccentricities of the inner and outer orbits, by leveraging the von Zeipel transformation technique. The resulting Hamiltonian model consists of three components: (a) the classical double-averaged quadrupole-order Hamiltonian, denoted by ${\cal F}_{20}$, (b) the classical Brown Hamiltonian, denoted by ${\cal F}_{21}$, which accounts for the nonlinear effects arising from short-period oscillations associated with the outer orbit, and (c) the extended Brown Hamiltonian, denoted by ${\cal F}_{22}$, representing the nonlinear effects from the short-period oscillations of the inner orbit. The introduction of the extended Brown Hamiltonian ${\cal F}_{22}$ into the standard long-term dynamical model is a key contribution of this work. Specifically, we define two factors, $\varepsilon_{21}$ and $\varepsilon_{22}$, to control the classical and extended Brown Hamiltonian, respectively. The first coefficient $\varepsilon_{21}$ (related to the single-averaging parameter $\epsilon_{\rm SA}$), standing for the ratio of the outer binary period to the ZLK timescale \citep{luo2016double}, is of order one with respect to the mean motion ratio $n_p/n$. The second coefficient $\varepsilon_{22}$, representing the ratio of the inner binary period to the ZLK timescale, is of order two in $n_p/n$. When the mean motion ratio is not small, the contribution from the extended Brown Hamiltonian becomes significant and must therefore be considered. For clarity, we refer to the complete model, comprising ${\cal F}_{20} + {\cal F}_{21} + {\cal F}_{22}$, as the extended Brown Hamiltonian model. 

The extended model exhibits the following features: (i) it is an integrable Hamiltonian system with a single degree of freedom, (ii) both $L = \sqrt{\mu a}$ and $H = G \cos{i}$ are conserved during long-term evolution, implying that the semimajor axis $a$ remains constant while the eccentricity $e$ exchanges with the inclination $i$, (iii) it is computationally efficient and readily applicable in practical scenarios, and (iv) it is of high accuracy in predicting long-term dynamical behavior, as demonstrated in the comparisons made in Figs. \ref{Fig2} and \ref{Fig3}.

Secondly, we provided a closed form for transformation between mean and osculating elements in terms of eccentricities of the inner and outer orbits, based on the generating function. This is the second contribution of this article. Furthermore, we have conducted numerical checks, demonstrating that our closed-form transformation is equivalent to the expansion-based transformation discussed in \citet{lei2019semi}.

The Hamiltonian framework developed in this study provides a fundamental dynamical model, which is particularly well-suited for studying von Zeipel–Lidov–Kozai oscillations in triple systems with low hierarchical configurations. Under this novel extended model, further investigations of modified ZLK dynamics and practical applications to irregular satellites of giant planets will be presented in subsequent papers in this series.

\section*{Acknowledgements}
Hanlun Lei would like to express gratitude to Prof. Scott Tremaine for helpful suggestions about the vectorial form of Brown Hamiltonian, to Prof. Xiyun Hou for valuable discussions about generating function in von Zeipel transformation, and to Dr. Hao Gao for his assistance in deriving the vectorial form of Brown corrections. In addition, we are grateful to the anonymous reviewer for his/her insightful comments, which help to improve the quality and readability of the manuscript. This work is financially supported by the National Natural Science Foundation of China (Nos. 12233003 and 12073011) and the China Manned Space Program with grant no. CMS-CSST-2025-A16. Evgeni Grishin acknowledges support from the ARC Discovery Program DP240103174 (PI: Heger).

\section*{Data Availability}
The original (osculating) orbit elements of Jupiter's irregular satellites can be extracted from Horizons system (https://ssd.jpl.nasa.gov/horizons/), where the reference frame is the ecliptic coordinate system of J2000.0 centered at Jupiter. The orbit elements are transformed to our defined coordinate system, and then transformation from osculating to mean elements is performed. Table \ref{Table1} shows the resulting osculating and mean elements of four example satellites at the epoch of October 18, 2024. In addition, the codes used in this article could be shared on reasonable request.



\bibliographystyle{mnras}
\bibliography{mybib} 




\appendix

\section{Expressions of unitary position vectors}\label{A1}
Under the coordinate system defined in this work, the unitary position vector of the perturber can be expressed in true anomaly $f_p$ by
\begin{equation}\label{EqA01}
{\hat{\bm r}_p} = \frac{{{{\bm r}_p}}}{{{r_p}}} = \left({\cos {f_p}},\;{\sin {f_p}},\;0 \right),
\end{equation}
and that of the inner test particle can be expressed in eccentric anomaly $E$ by
\begin{equation}\label{EqA02}
{\hat {\bm r}} = \frac{{{{\bm r}}}}{{{r}}}= \frac{1}{{1 - {e}\cos {E}}}\left[ \begin{array}{c}
{{ A}_1}\cos {E} + {{ B}_1}\sin {E} + {{ C}_1}\\
{{ A}_2}\cos {E} + {{ B}_2}\sin {E} + {{ C}_2}\\
{{ A}_3}\cos {E} + {{ B}_3}\sin {E} + {{ C}_3}
\end{array} \right],
\end{equation}
where the coefficients are
\begin{equation}\label{EqA03}
\begin{aligned}
{{A}_1} =& \cos{\Omega}\cos {\omega} - \cos {i} \sin {\Omega} \sin {\omega},\\
{{ B}_1} =&  - \eta \left( {\cos{\Omega}\sin {\omega} + \cos {i} \sin {\Omega} \cos {\omega}} \right),\; {{ C}_1} =  - {e}{{ A}_1},\\
{{ A}_2} =& \sin {\Omega}\cos {\omega} + \cos {i} \cos {\Omega} \sin {\omega},\\
{{ B}_2} =&  - \eta \left( {\sin {\Omega}\sin {\omega} - \cos {i} \cos {\Omega} \cos {\omega}} \right),\; {{ C}_2} =  - {e}{{ A}_2},\\
{{ A}_3} =& \sin {i}\sin {\omega},\; {{ B}_3} = \eta \sin {i}\cos {\omega},\; {{ C}_3} =  - {e}{{ A}_3}
\end{aligned}
\end{equation}
with $\eta=\sqrt{1-e^2}$. The same expressions about ${\hat{\bm r}_p}$ and ${\hat {\bm r}}$ can be found in \citet{lei2018modified}.

\section{Implementation of von Zeipel transformation}\label{A2}

The Hamiltonian can be decomposed into two parts,
\begin{equation}\label{EqA1}
{\cal H} = {{\cal H}_0} + \epsilon {{\cal H}_1},
\end{equation}
where the unperturbed term ${\cal H}_0$ governs the individual Keplerian motion of the test particle and perturber around the central body,
\begin{equation}\label{EqA2}
{{\cal H}_0} =  - \frac{{{\mu ^2}}}{{2{L^2}}} + {n_p}{L_p},
\end{equation}
and the perturbed term ${\cal H}_1$ stands for the third-body perturbation, given by
\begin{equation}\label{EqA3}
{{\cal H}_1} =  - \frac{{{\cal G}{m_p}}}{{{a_p}}}{\left( {\frac{a}{{{a_p}}}} \right)^2}{\left( {\frac{r}{a}} \right)^2}{\left( {\frac{{{a_p}}}{{{r_p}}}} \right)^3}\left( {\frac{3}{2}{{\cos }^2}\psi  - \frac{1}{2}} \right).
\end{equation}
In equation (\ref{EqA1}), $\epsilon$ is a formal parameter which is used to separate the perturbed Hamiltonian from the unperturbed one and it is taken as $\epsilon = 1$ in practical simulations \citep{beauge2006high}. Please bear in mind that the variables $a$, $r$, and $\cos {\psi}$ shown in ${{\cal H}_1}$ should be expressed in canonical variables defined by equation (\ref{Eq3}). Notice that the partition of Hamiltonian shown in equation (\ref{EqA1}) is similar to that of \citet{beauge2006high}, but is different from the form adopted in \citet{giacaglia1970semi} and later related studies. 

Considering the fact that there are two short-period angular coordinates ($l$ and $l_p$), we need to perform von Zeipel transformation twice: the first one is to remove short-period effects associated with $l$ and the next one is to remove short-period effects associated with $l_p$.

\textbf{Step I. Eliminate the angular coordinate $l (=M)$}

The purpose of this step is to make the transformed Hamiltonian be independent on the short-period angular coordinate $l$ through an identity transformation,
\begin{equation}\label{EqA4}
{\cal H}\left( {L,G,H,{L_p},l,g,h,{l_p}} \right) \equiv  {{\cal H}^*}\left( {L',G',H',{{L'_p}}, \sim ,g',h',{{l'_p}}} \right)
\end{equation}
where $\left( {L,G,H,{L_p},l,g,h,{l_p}} \right)$ is the set of canonical variables before transformation and $\left( {L',G',H',{{L'_p}}, l',g',h',{{l'_p}}} \right)$ is the corresponding set of variables after transformation. The canonical transformation between the old and new sets of variables is realized by
\begin{equation}\label{EqA5}
\begin{aligned}
L &= \frac{{\partial S}}{{\partial l}},\quad G = \frac{{\partial S}}{{\partial g}},\quad H = \frac{{\partial S}}{{\partial h}},\;{L_p} = \frac{{\partial S}}{{\partial {l_p}}},\\
l' &= \frac{{\partial S}}{{\partial L'}},\;\;g' = \frac{{\partial S}}{{\partial G'}},\;h' = \frac{{\partial S}}{{\partial H'}},\;{{l'_p}} = \frac{{\partial S}}{{\partial {{L'_p}}}},
\end{aligned}
\end{equation}
where $S$ is the generating function, given by
\begin{equation}\label{EqA6}
S\left( {L',G',H',{{L'_p}},l,g,h,{l_p}} \right) = {S_0} + \epsilon {S_1} + \epsilon^2 {S_2} +  \ldots
\end{equation}
with ${S_0} = L'l + G'g + H'h + {L'_p}{l_p}$ as the zero-order term. Performing Taylor expansion around $\left( {L',G',H',{{L'_p}},l,g,h,{l_p}} \right)$ for both sides of equation (\ref{EqA4}), we can get a series of homogeneous equations (up to order 2):
\begin{equation}\label{EqA7}
\begin{aligned}
\epsilon^0:\;{\cal H}_0^{\rm{*}} &= {{\cal H}_0},\\
\epsilon^1:\;{\cal H}_1^{\rm{*}} &= {{\cal H}_1} + \frac{{\partial {{\cal H}_0}}}{{\partial L'}}\frac{{\partial {S_1}}}{{\partial l}},\\
\epsilon^2:\;{\cal H}_2^{\rm{*}} &= \frac{{\partial {{\cal H}_0}}}{{\partial L'}}\frac{{\partial {S_2}}}{{\partial l}}{\rm{ + }}\frac{1}{2}\frac{{{\partial ^2}{{\cal H}_0}}}{{\partial {{L'}^{2}}}}{\left( {\frac{{\partial {S_1}}}{{\partial l}}} \right)^2} + \frac{{\partial {{\cal H}_1}}}{{\partial L'}}\frac{{\partial {S_1}}}{{\partial l}}\\
\;\;\;\;\;\;\; &+ \frac{{\partial {{\cal H}_1}}}{{\partial G'}}\frac{{\partial {S_1}}}{{\partial g}} + \frac{{\partial {{\cal H}_1}}}{{\partial H'}}\frac{{\partial {S_1}}}{{\partial h}} - \left( {\frac{{\partial {\cal H}_1^{{*}}}}{{\partial g}}\frac{{\partial {S_1}}}{{\partial G'}}{\rm{ + }}\frac{{\partial {\cal H}_1^{{*}}}}{{\partial h}}\frac{{\partial {S_1}}}{{\partial H'}}} \right).
\end{aligned}
\end{equation}
Solving the zeroth-order equation leads to
\begin{equation}\label{EqA8}
{\cal H}_0^{\rm{*}} =  - \frac{{{\mu ^2}}}{{2{{L'}^2}}} + {n_p}{L'_p}.
\end{equation}
According to the first-order perturbation equation, we can obtain the transformed Hamiltonian and the first-order generating function by
\begin{equation}\label{EqA9}
{\cal H}_1^{\rm{*}} = \frac{1}{{2\pi }}\int\limits_0^{2\pi } {{{\cal H}_1}{\rm d}l},\quad {S_1} = \frac{1}{n} \int {\left( {{\cal H}_1^{\rm{*}} - {{\cal H}_1}} \right){\rm d}l}.
\end{equation}
The transformed Hamiltonian ${\cal H}_1^{\rm{*}}$, corresponding to the so-called single-averaged Hamiltonian, is expressed as
\begin{equation}\label{EqA9-1}
\begin{aligned}
{\cal H}_1^{\rm{*}} &= \frac{3}{4}\frac{{G{m_p}}}{{{a_p}}}{\left( {\frac{a}{{{a_p}}}} \right)^2}{\left( {\frac{{{a_p}}}{{{r_p}}}} \right)^3}\left\{ {\frac{2}{3} + e^2} \right.\\
& - \left( {{A_1^2} + {B_1^2} + 2{C_1^2} - 2e{A_1}{C_1}} \right){\cos ^2}{f_p}\\
& - \left( {{A_2^2} + {B_2^2} + 2{C_2^2} - 2e{A_2}{C_2}} \right){\sin ^2}{f_p}\\
& - \left. {\left[ {{A_1}{A_2} + {B_1}{B_2} + 2{C_1}{C_2} - e\left( {{A_2}{C_1} + {A_1}{C_2}} \right)} \right]\sin \left( {2{f_p}} \right)} \right\}.
\end{aligned}
\end{equation}
and the associated generating function $S_1$, reflecting the short-period oscillations associated with the inner binary, is given by
\begin{equation}\label{EqA21}
\begin{aligned}
{S_1} &= \frac{{{{\cal C}_0}}}{n}\frac{{{{\left( {1 + {e_p}\cos {f_p}} \right)}^3}}}{{{{\left( {1 - e_p^2} \right)}^{3/2}}}}\left\{ {2{{\cal B}_p}\left[ {\left( {e{{\cal A}_p} - 4{{\cal C}_p}} \right)\cos E} \right.} \right.\\
&\left. { + \left( {e{{\cal C}_p} - {{\cal A}_p}} \right)\cos 2E + \frac{1}{3}e{{\cal A}_p}\cos 3E} \right]\\
& + \frac{1}{9}e\left[ {3\left( {{{\cal B}_p^2} - {{\cal A}_p^2}} \right) + {e^2}} \right]\sin 3E\\
& - \left[ {{{\cal B}_p^2} - {{\cal A}_p^2} + 2e{{\cal A}_p}{{\cal C}_p} + {e^2}} \right]\sin 2E\\
&\left. { + \left[ {e\left( {{{\cal B}_p^2} - {{\cal A}_p^2} - {e^2} + \frac{8}{3}} \right) + 4{{\cal A}_p}{{\cal C}_p}\left( {2 - {e^2}} \right)} \right]\sin E} \right\}
\end{aligned}
\end{equation}
with
\begin{equation}\label{EqA22}
\begin{aligned}
{{\cal A}_p} &= {A_1}\cos {f_p} + {A_2}\sin {f_p},\;{{\cal B}_p} = {B_1}\cos {f_p} + {B_2}\sin {f_p},\\
{{\cal C}_p} &= {C_1}\cos {f_p} + {C_2}\sin {f_p}.
\end{aligned}
\end{equation}

According to the second-order perturbation equation, we can obtain the transformed Hamiltonian by
\begin{equation}\label{EqA10}
\begin{aligned}
{\cal H}_2^{\rm{*}} &= \frac{1}{{2\pi }}\int\limits_0^{2\pi } {\left( {\frac{1}{2}\frac{{{\partial ^2}{{\cal H}_0}}}{{\partial {{L'}^{\;2}}}}{{\left( {\frac{{\partial {S_1}}}{{\partial l}}} \right)}^2} + \frac{{\partial {{\cal H}_1}}}{{\partial L'}}\frac{{\partial {S_1}}}{{\partial l}} + \frac{{\partial {{\cal H}_1}}}{{\partial G'}}\frac{{\partial {S_1}}}{{\partial g}}} \right){\rm d}l} \\
\;\;\;\;\; &+ \frac{1}{{2\pi }}\int\limits_0^{2\pi } {\left( {\frac{{\partial {{\cal H}_1}}}{{\partial H'}}\frac{{\partial {S_1}}}{{\partial h}} - \frac{{\partial {\cal H}_1^{\rm{*}}}}{{\partial g}}\frac{{\partial {S_1}}}{{\partial G'}} - \frac{{\partial {\cal H}_1^{\rm{*}}}}{{\partial h}}\frac{{\partial {S_1}}}{{\partial H'}}} \right){\rm d}l}. 
\end{aligned}
\end{equation}
The explicit expression of ${\cal H}_2^{\rm{*}}$ is not given here due to the space limitation. In order to derive a closed form of Hamiltonian and generating function expressed in terms of the eccentricity of the inner orbit ($e$), the following differential relation is used in equations (\ref{EqA9}) and (\ref{EqA10}),
\begin{equation}\label{EqA10-1}
{\rm d}l=(1-e\cos{E}){\rm d}E.
\end{equation}

\textbf{Step II. Eliminate the angular coordinate ${{l'_p (=M'_p)}}$}

A second von Zeipel transformation is performed in order to further remove the short-period angular coordinate ${{l'_p}}$ through an identity transformation,
\begin{equation}\label{EqA11}
\begin{aligned}
&{{\cal H}^*}\left( {L',G',H',{{L'_p}}, \sim ,g',h',{{l'_p}}} \right) \\
& \quad\quad\quad \equiv{{\cal H}^{*{\rm{*}}}}\left( {L'',G'',H'',{{L''_p}}, \sim ,g'',h'', \sim } \right)
\end{aligned}
\end{equation}
where $\left( {L',G',H',{{L'_p}}, l' ,g',h',{{l'_p}}} \right)$ is the set of variables before the second transformation and $\left( {L'',G'',H'',{{L''_p}}, l'' ,g'',h'',{{l''_p}}} \right)$ is the set of variables after the second transformation. Similarly, the generating function is denoted by
\begin{equation}\label{EqA12}
{S^{\rm{*}}} = L''l' + G''g' + H''h' + {L''_p}{l'_p} + \epsilon S_1^{\rm{*}} + \epsilon^2 S_2^{\rm{*}} +  \ldots 
\end{equation}
Following von Zeipel method, it is possible to arrive at a series of homogeneous equations up to order 2 as follows:
\begin{equation}\label{EqA13}
\begin{aligned}
\epsilon^0:\;{\cal H}_0^{{\rm{**}}} &= {\cal H}_0^*,\\
\epsilon^1:\;{\cal H}_1^{{\rm{**}}} &= {\cal H}_1^* + {n_p}\frac{{\partial S_1^*}}{{\partial {{l'_p}}}},\\
\epsilon^2:\;{\cal H}_2^{{\rm{**}}} &= {\cal H}_2^* + {n_p}\frac{{\partial S_2^*}}{{\partial {{l'_p}}}} + \frac{{\partial {\cal H}_1^*}}{{\partial G''}}\frac{{\partial S_1^*}}{{\partial g'}}\\
\;\;\;\;\; &+ \frac{{\partial {\cal H}_1^*}}{{\partial H''}}\frac{{\partial S_1^*}}{{\partial h'}} - \frac{{\partial {\cal H}_1^{{\rm{**}}}}}{{\partial g'}}\frac{{\partial S_1^*}}{{\partial G''}} - \frac{{\partial {\cal H}_1^{{\rm{**}}}}}{{\partial h'}}\frac{{\partial S_1^*}}{{\partial H''}}.
\end{aligned}
\end{equation}
Solving the zeroth-order equation leads to
\begin{equation}\label{EqA14}
{\cal H}_0^{{\rm{**}}} =  - \frac{{{\mu ^2}}}{{2{{L''}^2}}} + {n_p}{T''_p},
\end{equation}
which describes the Keplerian motion of the test particle and perturber around the central body. 

Solving the first-order perturbation equation, we can obtain the transformed Hamiltonian and the first-order generating function by
\begin{equation}\label{EqA15}
{\cal H}_1^{{\rm{**}}} = \frac{1}{{2\pi }}\int\limits_0^{2\pi } {{\cal H}_1^*{\rm d}{{l'_p}}},\quad S_1^* = \frac{1}{{{n_p}}}\int {\left( {{\cal H}_2^{{\rm{**}}} - {\cal H}_2^*} \right){\rm d}{{l'_p}}}.
\end{equation}
Here, the first-order transformed Hamiltonian ${\cal H}_1^{{\rm{**}}}$ corresponds to the classical quadrupole-order double-averaged Hamiltonian and it reads
\begin{equation}\label{EqA15-1}
{\cal H}_1^{{\rm{**}}} = - {\cal C}_0 {\cal F}_{20},
\end{equation}
where ${\cal F}_{20}$ is presented in equation (\ref{Eq6}) and the coefficient ${\cal C}_0$ is a constant during long-term evolution, given by
\begin{equation}\label{EqA15-2}
{{\cal C}_0} = \frac{3}{8}\frac{{{\cal G}{m_p}}}{{{a_p}}}{\left( {\frac{a}{{{a_p}}}} \right)^2}\frac{1}{{{{\left( {1 - e_p^2} \right)}^{3/2}}}}.
\end{equation}
The generating function $S_1^*$, reflecting the short-period oscillations associated with the inner binary, has the following form,
\begin{equation}\label{EqA23}
\begin{aligned}
S_1^* &=  - \frac{{{{\cal C}_0}}}{{{n_p}}}\left\{ {\left[ {{T_1} + {T_2} - 2e\left( {{T_4} + e} \right) - \frac{4}{3}} \right]\left( {{M_p} - {f_p}} \right)} \right.\\
& + \left( {{T_3} - e{T_5}} \right)\left[ {{e_p}\cos {f_p} + \cos 2{f_p} + \frac{1}{3}{e_p}\cos 3{f_p}} \right]\\
& - \frac{1}{2}{e_p}\left[ {3{T_1} + {T_2} - 2e\left( {{T_6} + 2e} \right) - \frac{8}{3}} \right]\sin {f_p}\\
&\left. { + \frac{1}{6}\left( {{T_2} - {T_1} + 2e{T_7}} \right)\left( {3\sin 2{f_p} + {e_p}\sin 3{f_p}} \right)} \right\}
\end{aligned}
\end{equation}
with
\begin{equation}\label{EqA24}
\begin{aligned}
{T_1} &= {A_1^2} + {B_1^2} + 2{C_1^2},\quad {T_2} = {A_2^2} + {B_2^2} + 2{C_2^2},\\
{T_3} &= {A_1}{A_2} + {B_1}{B_2} + 2{C_1}{C_2},\quad {T_4} = {A_1}{C_1} + {A_2}{C_2},\\
{T_5} &= {A_2}{C_1} + {A_1}{C_2},\quad {T_6} = 3{A_1}{C_1} + {A_2}{C_2},\\
{T_7} &= {A_1}{C_1} - {A_2}{C_2}.
\end{aligned}
\end{equation}
Solving the second-order perturbation equation, we can get the transformed Hamiltonian by computing the following definite integral,
\begin{equation}\label{EqA16}
{\cal H}_2^{{\rm{**}}} = \frac{1}{{2\pi }}\int\limits_0^{2\pi } {\left( {\frac{{\partial {\cal H}_1^*}}{{\partial G''}}\frac{{\partial S_1^*}}{{\partial g'}} + \frac{{\partial {\cal H}_1^*}}{{\partial H''}}\frac{{\partial S_1^*}}{{\partial h'}}} \right){\rm d}{{l'_p}}} + \frac{1}{{2\pi }}\int\limits_0^{2\pi } {{\cal H}_2^*{\rm d}{{l'_p}}}.
\end{equation}
If the transformation is implemented up to the second order, it is unnecessary to derive the second-order generating function $S_2^*$. About the quadratures in equations (\ref{EqA15}) and (\ref{EqA16}), the following differential relation is applied in order to reach a closed form in terms of the eccentricity of the outer orbit ($e_p$),
\begin{equation}\label{EqA17}
{\rm d}{{l'_p}}=\frac{(1-e_p^2)^{3/2}}{(1+e_p \cos{f'_p})^2}{\rm d}{f'_p}.
\end{equation}

In the right-hand side of equation (\ref{EqA16}), the first term corresponds to the classical Brown Hamiltonian term, coming from the accumulation of the short-period effects associated with the orbital motion of the outer binary (related to the generating function $S_1^*$), and the second term is the extended Brown Hamiltonian term, coming from the accumulation of the short-period effects associated with the orbital motion of the inner binary (related to the generating function $S_1$). In terms of mathematical expression, it reads 
\begin{equation}\label{EqA16-1}
{\cal H}_2^{{\rm{**}}} = - {\cal C}_0 \left(\varepsilon_{21} {\cal F}_{21} +\varepsilon_{22} {\cal F}_{22} \right),
\end{equation}
where ${\cal F}_{21}$ and ${\cal F}_{22}$ are, respectively, presented in equations (\ref{Eq7}) and (\ref{Eq8}). The coefficients $\varepsilon_{21}$ and $\varepsilon_{22}$ are, respectively, shown in equations (\ref{Eq9}) and (\ref{Eq10}).

After performing twice transformations, both the angular coordinates $l''$ and $l''_p$ are cyclic variables, leading to the fact that their conjugated momentum $L''$ and $T''_p$ become motion integral. Consequently, in the long-term evolution ${\cal H}_0^{{\rm{**}}}$ remains stationary and can be removed from the Hamiltonian. In addition, the gauge freedom discussed in \citet{tremaine2023hamiltonian} is taken into consideration in order to further remove the longitude of ascending node from the resulting Hamiltonian, making the angular momentum along the $z$-axis be conserved.

To summarize, the closed form of Hamiltonian function after removing $l$ and $l_p$ is (here the zeroth-order constant term is ignored)
\begin{equation}\label{EqA18}
\begin{aligned}
{{\cal H}^{**}} &= \epsilon {\cal H}_1^{**} + \epsilon^2 {\cal H}_2^{**} \\
&=  - {{\cal C}_0}{\cal F} = - {\cal C}_0 \left({\cal F}_{20} + \varepsilon_{21} {\cal F}_{21} +\varepsilon_{22} {\cal F}_{22} \right),
\end{aligned}
\end{equation}
where the expression of ${\cal F}$ has been presented in equation (\ref{Eq5}) and the coefficient ${\cal C}_0$ can be found in equation (\ref{EqA15-2}). The closed form of first-order generating function is written as
\begin{equation}\label{EqA20}
S = {S_1} + S_1^*,
\end{equation}
where ${S_1}$ is the generating function of \textbf{Step I} given by equation (\ref{EqA21}), and $S_1^*$ is the generating function of \textbf{Step II} given by equation (\ref{EqA23}). Notice that the generating function shown here can be used to realize transformations between mean and osculating elements. Please see Section \ref{Sect4} for more details.

\section{Equations of motion}\label{A4}

Under the extended Brown Hamiltonian model, the evolution equations of eccentricity, argument of pericenter, and longitude of ascending node are
\begin{equation}\label{EqA4-1}
\begin{aligned}
\frac{{{\rm d}e}}{{{\rm d}t}} &=  - \frac{\eta }{{n{a^2}e}}\frac{{\partial {\cal F}}}{{\partial \omega }},\\
\frac{{{\rm d}\omega }}{{{\rm d}t}} &= \frac{\eta }{{n{a^2}}}\left( {\frac{1}{e}\frac{{\partial {\cal F}}}{{\partial e}} - \frac{{\cot i}}{{{\eta ^2}}}\frac{{\partial {\cal F}}}{{\partial i}}} \right),\\
\frac{{{\rm d}\Omega }}{{{\rm d}t}} &= \frac{1}{{n{a^2}\eta }}\frac{1}{{\sin i}}\frac{{\partial {\cal F}}}{{\partial i}},\\
\end{aligned}
\end{equation}
where ${\cal F}$ is given by equation (\ref{Eq5}) and $\eta = \sqrt{1-e^2}$. It should be noted that the inclination $i$ can be derived from the motion integral $H = \sqrt{1-e^2}\cos{i}$, which is determined by the initial condition. For convenience, let us denote the equations of motion as follows:
\begin{equation}\label{EqA4-2}
\begin{aligned}
\frac{{{\rm d}e}}{{{\rm d}t}} &= {\left( {\frac{{{\rm d}e}}{{{\rm d}t}}} \right)_{20}} + {\varepsilon _{21}}{\left( {\frac{{{\rm d}e}}{{{\rm d}t}}} \right)_{21}} + {\varepsilon _{22}}{\left( {\frac{{{\rm d}e}}{{{\rm d}t}}} \right)_{22}},\\
\frac{{{\rm d}\omega }}{{{\rm d}t}} &= {\left( {\frac{{{\rm d}\omega }}{{{\rm d}t}}} \right)_{20}} + {\varepsilon _{21}}{\left( {\frac{{{\rm d}\omega }}{{{\rm d}t}}} \right)_{21}} + {\varepsilon _{22}}{\left( {\frac{{{\rm d}\omega }}{{{\rm d}t}}} \right)_{22}},\\
\frac{{{\rm d}\Omega }}{{{\rm d}t}} &= {\left( {\frac{{{\rm d}\Omega }}{{{\rm d}t}}} \right)_{20}} + {\varepsilon _{21}}{\left( {\frac{{{\rm d}\Omega }}{{{\rm d}t}}} \right)_{21}} + {\varepsilon _{22}}{\left( {\frac{{{\rm d}\Omega }}{{{\rm d}t}}} \right)_{22}},
\end{aligned}
\end{equation}
where the contribution from the double-averaged quadrupole-order term is
\begin{equation}\label{EqA4-3}
\begin{aligned}
{\left( {\frac{{{\rm d}e}}{{{\rm d}t}}} \right)_{20}} &= \frac{{5\eta }}{{n{a^2}}} e {\sin ^2}i\sin 2\omega, \\
{\left( {\frac{{{\rm d}\omega }}{{{\rm d}t}}} \right)_{20}} &= - \frac{{  1}}{{n{a^2}\eta }}\left[ {{\eta ^2}\left( {1 - 5\cos 2\omega } \right) - 10{{\cos }^2}i{{\sin }^2}\omega } \right],\\
{\left( {\frac{{{\rm d}\Omega }}{{{\rm d}t}}} \right)_{20}} &=  - \frac{1}{{n{a^2}\eta }} {\cos i} \left[ {2 + {e^2}\left( {3 - 5\cos 2\omega } \right)} \right],
\end{aligned}
\end{equation}
the contribution from the classical Brown correction is
\begin{equation}\label{EqA4-4}
\begin{aligned}
{\left( {\frac{{{\rm d}e}}{{{\rm d}t}}} \right)_{21}} &= \frac{{45{\eta ^2}}}{{8n{a^2}}}e\cos i{\sin ^2}i\sin 2\omega, \\
{\left( {\frac{{{\rm d}\omega }}{{{\rm d}t}}} \right)_{21}} &= \frac{{9}}{{8n{a^2}}} {\cos i} \left[ {{\eta ^2}\left( {11 + 5\cos 2\omega } \right) + 10{{\cos }^2}i{{\sin }^2}\omega } \right],\\
{\left( {\frac{{{\rm d}\Omega }}{{{\rm d}t}}} \right)_{21}} &=  - \frac{3}{{16n{a^2}}}\left[ {2 + {e^2}\left( {33 + 15\cos 2\omega } \right)} \right.\\
& - \left. {  {{\cos }^2}i\left( {6 - 51{e^2} + 45{e^2}\cos 2\omega } \right)} \right],
\end{aligned}
\end{equation}
and the contribution from the extended Brown correction is
\begin{equation}\label{EqA4-5}
\begin{aligned}
{\left( {\frac{{{\rm d}e}}{{{\rm d}t}}} \right)_{22}} &= \frac{{e\eta {{\sin }^2}i}}{{128{a^2}n}}\left\{ { - 285{e^2}{{\sin }^2}i\sin 4\omega } \right.\\
& + \left. {  \left[ {1386 - 337{e^2} + \left( {270 - 555{e^2}} \right)\cos 2i} \right]\sin 2\omega } \right\} ,\\
{\left( {\frac{{{\rm d}\omega }}{{{\rm d}t}}} \right)_{22}} &= \frac{1}{{128n{a^2}\eta }}\left\{ {{\eta ^2}\left[ {681\left( {4 + {e^2}} \right) + 4\left( {279 + 109{e^2}} \right)\cos 2\omega } \right.} \right.\\
& - \left. {  285{e^2}\cos 4\omega } \right] + {\cos ^2}i\left[ {704 + 285{e^2}\left( {2 - {e^2}} \right)\cos 4\omega } \right.\\
& - \left. {  16\left( {36 + 166{e^2} - 83{e^4}} \right)\cos 2\omega  + 1830{e^2} - 915{e^4}} \right]\\
& + \left. {  60{{\cos }^4}i\left( {18 - 37{e^2} + 19{e^2}\cos 2\omega } \right){{\sin }^2}\omega } \right\},\\
{\left( {\frac{{{\rm d}\Omega }}{{{\rm d}t}}} \right)_{22}} &= \frac{{\cos i}}{{256n{a^2}\eta }}\left\{ {920 - 3576{e^2} + 273{e^4}} \right.\\
& + 3\left( {56 - 472{e^2} + 701{e^4}} \right)\cos 2i - 570{e^4}{\sin ^2}i\cos 4\omega \\
&+ \left. {  4{e^2}\left[ {558 + 109{e^2} + \left( {270 - 555{e^2}} \right)\cos 2i} \right]\cos 2\omega } \right\}.
\end{aligned}
\end{equation}

\section{Vectorial form of the extended Brown Hamiltonian}
\label{A5}
In terms of the normalized angular momentum vector ${\bm j} = \left(j_x,j_y,j_z\right)$ and the eccentricity vector ${\bm e} = \left(e_x,e_y,e_z\right)$, the classical double-averaged quadrupole-order Hamiltonian can be written as
\begin{equation}\label{EqA5-1}
{{\cal F}_{20}} =  2{e^2}- 5{e_z^2}  + {j_z^2} - \frac{1}{3},
\end{equation}
the classical Brown Hamiltonian correction becomes
\begin{equation}\label{EqA5-2}
{{\cal F}_{21}} = \frac{3}{8}{j_z}\left( {1 - j_z^2 + 24{e^2} - 15e_z^2} \right),
\end{equation}
and the extended Brown Hamiltonian correction is
\begin{equation}\label{EqA5-3}
\begin{aligned}
{{\cal F}_{22}} &= \frac{1}{{64}}\left\{{ 8e^2\left(13{e^2} + {22e_z^2 + 4j_z^2 + 120}\right) - 94j_z^2} \right.\\
&\left. { - 3\left[ {95e_z^4 + 6e_z^2\left( {15j_z^2 + 31} \right) + 7j_z^4} \right]} \right\},
\end{aligned}
\end{equation}
where the normalized angular momentum vector ${\bm j}$ and the eccentricity vector ${\bm e}$ are given by \citep{tremaine2023hamiltonian}
\begin{equation}\label{EqA5-4}
{\bm j} = \left( {\begin{array}{*{20}{c}}
{{j_x}}\\
{{j_y}}\\
{{j_z}}
\end{array}} \right) = \sqrt {1 - {e^2}} \left( {\begin{array}{*{20}{c}}
{\sin i\sin \Omega }\\
{ - \sin i\cos \Omega }\\
{\cos i}
\end{array}} \right),
\end{equation}
and
\begin{equation}\label{EqA5-5}
{\bm e} = \left( {\begin{array}{*{20}{c}}
{{e_x}}\\
{{e_y}}\\
{{e_z}}
\end{array}} \right) = e\left( {\begin{array}{*{20}{c}}
{\cos \Omega \cos \omega  - \cos i\sin \Omega \sin \omega }\\
{\sin \Omega \cos \omega  + \cos i\cos \Omega \sin \omega }\\
{\sin i\sin \omega }
\end{array}} \right).
\end{equation}

\section{Octupole-level Hamiltonian}
\label{A3}
\begin{figure}
\centering
\includegraphics[width=\columnwidth]{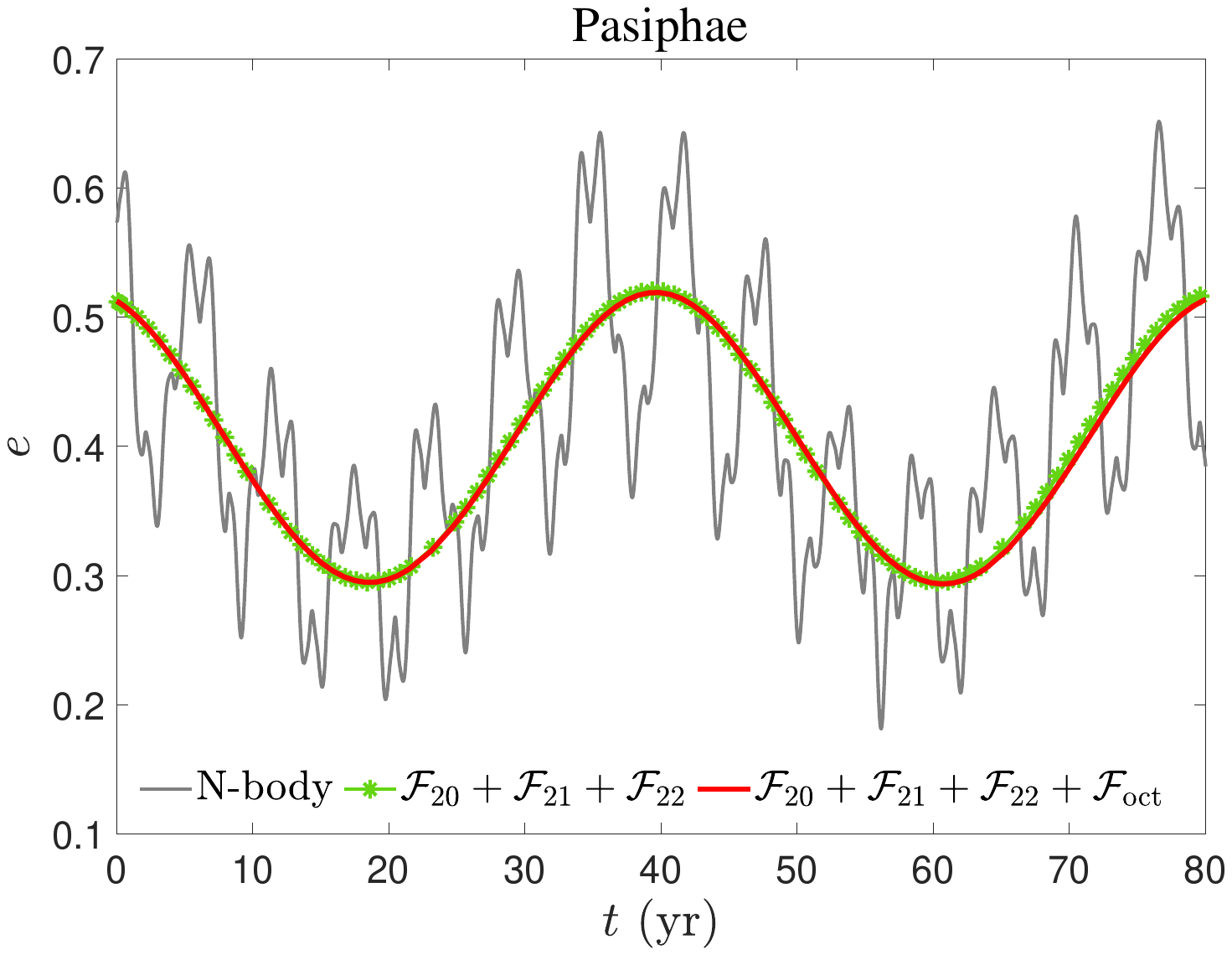}
\caption{Eccentricity evolutions of Pasiphae under the extended Brown Hamiltonian models with and without octupole-level term. The trajectories propagated under the full $N$-body model are shown in grey lines as background.}
\label{FigA1}
\end{figure}

When the octupole-order term is included, the complete Hamiltonian (normalized by the constant coefficient ${\cal C}_0$) can be written as
\begin{equation}\label{EqA24-1}
{\cal F} = \left({\cal F}_{20} + \varepsilon_{21} {\cal F}_{21} + \varepsilon_{22}{\cal F}_{22}\right) + \varepsilon_{\rm oct}{\cal F}_{\rm oct}
\end{equation}
where the terms in bracket stand for the (linear and nonlinear) quadrupole-level Hamiltonian which has been discussed in the main body of this paper (see equation \ref{Eq5}) and $\varepsilon_{\rm oct} {\cal F}_{\rm oct}$ stands for the octupole-level contribution, given by \citep{li2014,tremaine2023hamiltonian,naoz2016eccentric,lithwick2011eccentric,ford2000secular,katz2011long,klein2024hierarchical}
\begin{equation}\label{EqA24-2}
\varepsilon_{\rm oct}  = \left(\frac{a}{{{a_p}}}\right)\frac{{{e_p}}}{{1 - e_p^2}},
\end{equation}
and
\begin{equation}\label{EqA24-3}
\begin{aligned}
{{\cal F}_{\rm oct}} &= \frac{5}{{16}}\left( {e + \frac{3}{4}{e^3}} \right) \left[ {D_1\cos \left( {\omega  - \Omega } \right)} + D_2\cos \left( {\omega  + \Omega } \right) \right]\\
& - \frac{{175}}{{64}}{e^3}\left[ {D_3\cos \left( {3\omega  - \Omega } \right)} { + D_4 \cos \left( {3\omega  + \Omega } \right)} \right],
\end{aligned}
\end{equation}
where
\begin{equation*}
\begin{aligned}
D_1 &=  {1 - 11\cos i - 5{{\cos }^2}i + 15{{\cos }^3}i},\\
D_2 &=  {1 + 11\cos i - 5{{\cos }^2}i - 15{{\cos }^3}i},\\
D_3 &=  {1 - \cos i - {{\cos }^2}i + {{\cos }^3}i},\\
D_4 &=  {1 + \cos i - {{\cos }^2}i - {{\cos }^3}i}.
\end{aligned}
\end{equation*}
In particular, the octupole-order Hamiltonian vanishes when the perturber is moving on a circular orbit. For Jupiter's irregular satellite Pasiphae, its octupole-related coefficient is $\varepsilon_{\rm oct} = 1.45 \times 10^{-3}$. Figure \ref{FigA1} presents its eccentricity evolutions under the extended models with and without octupole-level term, together with the $N$-body simulations as the background. It is observed that, for the considered example, the inclusion of octupole-order term has negligible contribution to the ZLK oscillation.

\bsp	
\label{lastpage}
\end{document}